\documentclass[a4paper,fleqn,usenatbib,useAMS]{mnras}

\usepackage{amssymb}	% Extra maths symbols
\usepackage{multicol}        % Multi-column entries in tables
\usepackage{bm}		% Bold maths symbols, including upright Greek
\usepackage{pdflscape}	% Landscape pages
\usepackage[T1]{fontenc}
\usepackage{ae,aecompl}

\usepackage {graphicx}
\usepackage{natbib}
\usepackage{aas_journals}
\usepackage{color}
\usepackage{amsmath}
\definecolor{mygray}{gray}{0.7}

\def\aap{{ A\&A}}

\def\apjl{{ApJL}}

\def\apjs{{ApJS}}

\def\redmagic{redMaGiC}

\newcommand{\bes}{\begin{equation*}}
\newcommand{\ees}{\end{equation*}}
\newcommand{\bea}{\begin{eqnarray}}
\newcommand{\eea}{\end{eqnarray}}
\newcommand{\beas}{\begin{eqnarray*}}
\newcommand{\eeas}{\end{eqnarray*}}

\newcommand{\ltsima}{$\; \buildrel < \over \sim \;$}
\newcommand{\lsim}{\lower.5ex\hbox{\ltsima}}
\newcommand{\gtsima}{$\; \buildrel > \over \sim \;$}
\newcommand{\gsim}{\lower.5ex\hbox{\gtsima}}

\def\gtrsim{\mathrel{\hbox{\rlap{\hbox{\lower4pt\hbox{$\sim$}}}\hbox{$>$}}}}
\let\ga=\gtrsim
\def\lesssim{\mathrel{\hbox{\rlap{\hbox{\lower4pt\hbox{$\sim$}}}\hbox{$<$}}}}
\let\la=\lesssim

\definecolor{mygray}{gray}{0.5}

\newcommand{\be}{\begin{equation}}
\newcommand{\ee}{\end{equation}}
\newcommand{\ba}{\begin{eqnarray}}
\newcommand{\ea}{\end{eqnarray}}

\title[Imprint of DES super-structures on the CMB]{Imprint of DES super-structures on the Cosmic Microwave Background}
\author[Kov\'{a}cs et al.]{
\parbox{\textwidth}{
\LARGE
A.~Kov\'{a}cs$^1$\thanks{Corresponding author: \texttt{\rm \texttt{akovacs@ifae.es}}},
C.~S\'{a}nchez$^{1}$,
J.~Garc\'ia-Bellido$^{2}$,
S.~Nadathur$^{3}$,
R.~Crittenden$^{3}$,
D.~Gruen$^{4,5,6}$,
D.~Huterer$^{7}$,
D.~Bacon$^{3}$,
J.~DeRose$^{8,4}$,
S.~Dodelson$^{9,10}$,
E.~Gazta\~{n}aga$^{11}$,
D.~Kirk$^{12}$,
O.~Lahav$^{12}$,
R.~Miquel$^{1,13}$,
K.~Naidoo$^{12}$,
J.~A.~Peacock$^{38}$
B.~Soergel$^{14,15}$,
L.~Whiteway$^{12}$,
F.~B.~Abdalla$^{12,16}$,
S.~Allam$^{9}$,
J.~Annis$^{9}$,
A.~Benoit-L{\'e}vy$^{17,11,18}$,
E.~Bertin$^{17,18}$,
D.~Brooks$^{7}$,
E.~Buckley-Geer$^{9}$,
A. Carnero Rosell$^{19,20}$,
M.~Carrasco~Kind$^{21,22}$,
J.~Carretero$^{11,1}$,
C.~E.~Cunha$^{4}$,
C.~B.~D'Andrea$^{3,23}$,
L.~N.~da Costa$^{19,20}$,
D.~L.~DePoy$^{24}$,
S.~Desai$^{25,26}$,
T.~F.~Eifler$^{27}$,
D.~A.~Finley$^{9}$,
B.~Flaugher$^{9}$,
P.~Fosalba$^{11}$,
J.~Frieman$^{9,10}$,
T. ~Giannantonio$^{14,15}$,
D.~A.~Goldstein$^{28,29}$,
R.~A.~Gruendl$^{21,22}$,
G.~Gutierrez$^{9}$,
D.~J.~James$^{30}$,
K.~Kuehn$^{31}$,
N.~Kuropatkin$^{9}$,
J.~L.~Marshall$^{32}$,
P.~Melchior$^{33}$,
F. ~Menanteau$^{21,22}$,
B.~Nord$^{9}$,
R. Ogando$^{19,20}$,
A.~A.~Plazas$^{27}$,
A.~K.~Romer$^{34}$,
E.~Sanchez$^{35}$,
V.~ Scarpine$^{9}$,
I.~Sevilla-Noarbe$^{35}$,
F.~Sobreira$^{36,19}$,
E.~Suchyta$^{37}$,
M.~Swanson$^{22}$,
G.~Tarle$^{7}$,
D.~Thomas$^{3}$,
A.~R.~Walker$^{30}$}
  \vspace{0.125cm}\\~\\
\parbox{\textwidth}{\centering \textsc{\Large(The DES Collaboration)} \\ \centering \textit{Author affiliations are listed at the end of this paper}\\ }}

\begin{document}
\date{Submitted 2016}
\pagerange{\pageref{firstpage}--\pageref{lastpage}} \pubyear{2016}
\maketitle
\label{firstpage}
\begin{abstract}
Small temperature anisotropies in the Cosmic Microwave Background can be sourced by density perturbations via the late-time integrated Sachs-Wolfe effect.
Large voids and superclusters are excellent environments to make a localized measurement of this tiny imprint. In some cases excess signals have been reported. We probed these claims with an independent data set, using the first year data of the Dark Energy Survey in a different footprint, and using a different super-structure finding strategy. We identified 52 large voids and 102 superclusters at redshifts $0.2 < z < 0.65$. We used the Jubilee simulation to {\it a priori} evaluate the optimal ISW measurement configuration for our compensated top-hat filtering technique, and then performed a stacking measurement of the CMB temperature field based on the DES data. For optimal configurations, we detected a cumulative cold imprint of voids with $\Delta T_{f} \approx -5.0\pm3.7~\mu K$ and a hot imprint of superclusters $\Delta T_{f} \approx 5.1\pm3.2~\mu K$ ; this is $\sim1.2\sigma$ higher than the expected $|\Delta T_{f}| \approx 0.6~\mu K$ imprint of such super-structures in $\Lambda$CDM. If we instead use an {\it a posteriori} selected filter size ($R/R_{v}=0.6$), we can find a temperature decrement as large as $\Delta T_{f} \approx -9.8\pm4.7~\mu K$ for voids, which is $\sim2\sigma$ above $\Lambda$CDM expectations and is comparable to previous measurements made using SDSS super-structure data.
\end{abstract}
\section{Introduction}

The largest observable structures in the low-redshift Universe leave their mark on the Cosmic Microwave Background (CMB) radiation, directly probing the physics of Dark Energy. The physical mechanism by which large voids and superclusters induce secondary anisotropies in the CMB to the primary fluctuations of the CMB is called the Integrated Sachs-Wolfe effect \citep[ISW]{SachsWolfe} in the linear regime, and the Rees-Sciama effect \citep[RS]{ReesSciama} on smaller scales.

In the concordance $\Lambda$CDM framework, the maximum unfiltered ISW imprint in the centre of typical (and thus numerous) voids and superclusters is of the order of $|\Delta T_{c} | \leq 1 ~\mu K$, and can grow to $|\Delta T_{c} | \approx 20 ~\mu K$ for the largest and rarest observable super-structures \citep{SzapudiEtAl2014,Nadathur2014,Sahlen2015}. Using a compensated top-hat (CTH) filter reduces the signal, with $|\Delta T_{f}|\sim |\Delta T_{c} |/2$ at best. The non-linear RS effects remain subdominant, contributing at most $\sim 10 \%$  of the linear ISW signal on small scales and higher redshifts \citep{CaiEtAl2010}; however their magnitude and relative strength may be different in alternative cosmological models \citep{CaiEtAl2014}. Measuring the ISW and RS imprints of typical super-structures is a challenging task in the presence of the strong primordial CMB temperature fluctuations that are effectively noise in this case \citep[e.g.][]{b11}. 

Traditionally, the weak ISW signal is measured in the angular cross-correlation of galaxy density maps and the CMB temperature field, leading to marginally and moderately significant detections \citep[e.g.][]{Fosalba2003,Fosalba2004,ho,gianEtAl08,b14,gian,KovacsEtAl2013,Planck19,PlackISW2015}. However, \cite{GranettEtAl2008} (Gr08, hereafter) concentrated instead on mapping large-scale peaks and troughs in the galaxy density field, where the ISW effect is expected to be the strongest; they used the \texttt{ZOBOV} algorithm \citep{ZOBOV} to obtain a catalog of significant supervoids and superclusters using the Sloan Digital Sky Survey (SDSS) Data Release 4 (DR4) Mega-z photometric LRG catalog \citep{megaz} with some additional data based on DR6 photometric redshifts. The super-structure locations were then used for stacking the CMB temperature maps, using a CTH filter. This simple filtering statistic averages the $\Delta T$ CMB temperatures centred on the structures within a circular aperture $r < R$ for filter size $R$, from which the background temperature is subtracted over a concentric equal-area annulus, $R < r < \sqrt{2} R$.

Using those SDSS supervoids and superclusters seen to be the most probable (i.e., least likely to occur in random catalogues), Gr08 found $\Delta T_{f} = -11.3 \pm 3.1 ~\mu K$ and $\Delta T_{f} = 7.9 \pm 3.1 ~\mu K$, respectively, using a fixed aperture size of $R=4^{\circ}$. The combined $|\Delta T_{f}| = 9.6 \pm 2.2 ~\mu K$ signal appears to be $\gsim3\sigma$ higher than $\Lambda$CDM expectations, according to theoretical and simulated follow-up studies \citep{PapaiEtAl2011,PapaiSzapudi2010,Nadathur2012,Flender2013,CaiEtAl2014,Hotchkiss2015,Aiola}. Notably, \cite{Hernandez2013} found that varying the number of the objects in the stacking, or using different filter sizes typically lowers the overall significance. Otherwise the original Gr08 signal has survived new CMB data releases and tests against CMB and galactic systematics and remains a puzzle. 

Recently, several CMB stacking analyses based on the same filtering technique have been performed using other void and supercluster catalogues drawn from galaxy samples with spectroscopic redshifts \citep{Ilic2013,CaiEtAl2014,Planck19,Hotchkiss2015,Cai2016}. No high-significance detection comparable to that of Gr08 has been observed, although \cite{CaiEtAl2014} and \cite{Cai2016} report marginally significant (at $\lesssim 2\sigma$) detections of a correlation, with amplitude still exceeding $\Lambda$CDM expectations. Using a different technique based on optimal matched filters, \cite{NadathurCrittenden2016} reported a significant detection of the ISW signal from voids and superclusters, but in this case with amplitude consistent with $\Lambda$CDM.

The Mega-z LRG tracer catalogue used by Gr08 used photometric redshifts which smear the galaxy distribution along the line-of-sight (LOS, hereafter); 
this could potentially lead to biases that have not been studied in detail using simulations or accounted for in modelling the ISW effect of voids. \cite{Granett2015} recently reconstructed the average shape of the Gr08 supervoids using a BOSS DR12 galaxy catalogue, and found that the supervoids are significantly elongated in the LOS with an axis ratio $R_{\parallel}/R_{\perp} \approx 2.6\pm 0.4$ based on estimates of the stacked LOS ($R_{\parallel}$) and the transverse ($R_{\perp}$) radii of the supervoids. No evidence for a significant LOS elongation was found for the Gr08 supercluster sample.

This elongation of structures or considerations of multiple voids in alignment \citep{Naidoo2016} might shed new light on the Gr08 measurement, as ISW-RS expectations for prolate supervoids should be higher than in the spherical case \citep{MarcosCaballero2015}. The significant LOS elongation of the Eridanus supervoid \citep{SzapudiEtAl2014}, reported by \cite{KovacsJGB2015}, also suggests stronger contributions to the Cold Spot via ISW-RS effects than expected previously. These findings motivate further studies of the ISW imprints of large voids and superclusters, especially using photometric redshift surveys that densely sample large physical volumes.

In this paper, we used novel algorithms developed by \cite{Sanchez2016} based on the void finder presented in \cite{ClampittJain2015}. Identifying voids in photometric data is non-trivial and requires special techniques. However, \cite{Sanchez2016} measured the weak lensing effects of voids identified the Science Verification data of the Dark Energy Survey \citep[DES,][]{DES} and proved that their voids are truly underdense in the matter field. We now extended this void finding procedure to a larger DES data set using the first year of observations. We also inverted this void finder algorithm to detect extended overdensities (superclusters), and tested the possible elongation of our super-structures in DES mock galaxy catalogues. We then measured the expected ISW imprint of voids and superclusters using the Jubilee simulation\footnote{http://jubilee.ft.uam.es} and its corresponding ISW map \citep{Watson2014}. This analysis serves as a test case where we know that super-structures leave an imprint in the projected ISW-only map. Our goal was to characterize the shape and amplitude of the imprints in the simulation and then perform the measurements with DES data using {\it a priori} selected measurement parameters. 

The paper is organized as follows. Data sets, algorithms, and super-structure properties are introduced in Section 2. Our simulation analyses are presented in Section 3, while Section 4 introduces our observational results. The final section contains a summary, discussion and interpretation of our findings. 

\section{Data sets for the ISW analysis}

\subsection{CMB data}
We used {\it Planck}'s SMICA map \citep{Planck_15} downgraded to $N_{\rm side}=512$ resolution with \texttt{HEALPix}  pixelization \citep{healpix}. SMICA produces CMB maps by linearly combining all {\it Planck} input channels with multipole-dependent weights, including multipoles up to $\ell<4000$. Potentially contaminated CMB pixels with high Galactic dust or at locations of known point sources were masked out based on the $N_{\rm side}=512$ WMAP 9-year extended temperature analysis mask \citep{WMAP9} to avoid repixelization effects of the $N_{\rm side}=2048$ CMB masks provided by {\it Planck}. It has already been pointed out by \cite{GranettEtAl2008}, and later confirmed by \cite{Ilic2013}, \cite{Planck19}, and \cite{CaiEtAl2013} that the ISW-like cross-correlation signal detected at void locations is independent of the CMB data set when looking at WMAP Q, V, W, or {\it Planck} temperature maps. We thus limited our analysis to the latest {\it Planck} SMICA sky map.

\subsection{The DES \redmagic\ catalog}

The Dark Energy Survey is a photometric redshift survey that will cover about one eighth of the sky (5000 sq.~deg.) to a depth of $i_{AB} < 24$, imaging about 300 million galaxies in 5 broadband filters ($grizY$) up to redshift $z=1.4$ \citep{DECam,morethanDE2016}.

In this paper we used a luminous red galaxy sample from the first year of observations (Y1A1). This red-sequence Matched-filter Galaxy Catalog \cite[\redmagic,][]{Rozo2015} is a catalog of photometrically selected luminous red galaxies, based on the red-sequence matched-filter Probabalistic Percolation (redMaPPer) cluster finder algorithm \citep{Rykoff2014}. Specifically, \redmagic\ uses the redMaPPer-calibrated model for the color of red-sequence galaxies as a function of magnitude and redshift. This model is used to find the best fit photometric redshift for all galaxies irrespective of type, and the $\chi^2$ goodness-of-fit of the model is computed. For each redshift slice, all galaxies fainter than some minimum luminosity threshold $L_{\rm min}$ are rejected. In addition, \redmagic\ applies a $\chi^2$ cut $\chi^2 \leq \chi_{\rm max}^2$, where the cut $\chi_{\rm max}^2$ as a function of redshift is chosen to ensure that the resulting galaxy sample has a constant comoving space density in two versions; $\bar{n}\approx2\times10^{-4}h^{3}$ Mpc$^{-3}$ (high luminosity sample) and $\bar{n}\approx 10^{-3}h^{3}$ Mpc$^{-3}$ (high density sample).

The luminosity cut is $L\geq L_*(z)$ and $L\geq L_*(z)/2$ for the high luminosity and high density samples, respecitvely, where the value of $L_*(z)$ at {\it z=0.1} is set to match the redMaPPer definition for SDSS, and the redshift evolution for $L_*(z)$ is that predicted using a simple passive evolution starburst model at $z=3$.

We utilized the \redmagic\ sample because of the exquisite photometric redshifts of the \redmagic\ galaxy catalog, namely $\sigma_z/(1+z)\approx 0.02$, and a $4\sigma$ redshift outlier rate of $r_\mathrm{out}\simeq1.41\%$. For DES main galaxies, a significantly larger $\sigma_z/(1+z)\approx 0.1$ typical photo-$z$ error has been estimated by \cite{Sanchez2014}, corresponding to $\sim220~h^{-1}{\rm Mpc}$ at $z\approx0.6$. Also, since void properties depend on the tracer sample used, the constant density of \redmagic\ tracers helps in assuring the resulting voids have similar properties \citep{Sanchez2016}. A redshift-independent linear galaxy bias of $b=1.6$ was assumed by \cite{Gruen2016} for this data set in a similar DES analysis.

We restricted our analysis to a rectangular area at $5^{\circ}<RA<100^{\circ}$ and $-58^{\circ}<Dec<-42^{\circ}$ inside the largest contiguous patch of the Y1 footprint, as shown in Figure 1.

\begin{figure}
\begin{center}
\includegraphics[width=85mm]{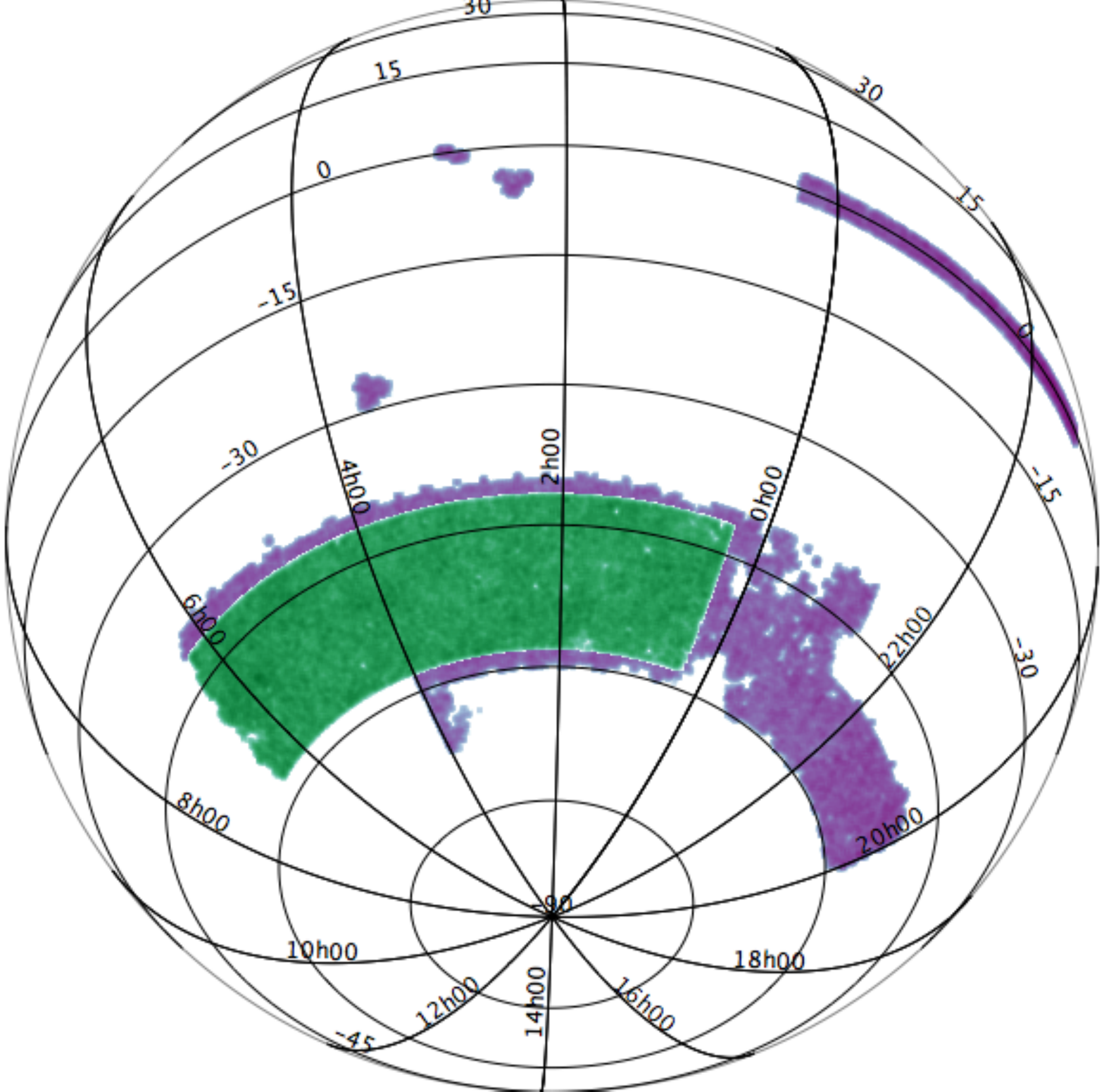}
\label{footprint}
\caption{DES Y1 survey footprint in purple and the rectangular area used for void finding in green. We focused on inner areas in the Y1 footprint without significant holes and complicated mask features to ensure the accuracy of the void finder. This is the same for the real data and mock DES data.}
\end{center}
\end{figure}

We also relied on Y1A1 Buzzard redMaGiC simulations for validating our super-structure catalogues. Photo-$z$ characteristics, sample density, and the sky coverage are identical to those of the real data set for this realistic mock galaxy catalogue. We used the official DES Y1A1 \redmagic\ mask.

\subsection{A catalog of super-structures in DES}

We identified voids in Y1A1 redMaGiC galaxy data and simulations using the void finder tool described in \cite{Sanchez2016}. The heart of the method is a restriction to 2D slices of galaxy data, and measurements of the projected density field around centers defined by minima in the corresponding smoothed density field.

Analyses of realistic DES \redmagic\ simulations confirm that significant real underdensities can be identified in slices of width roughly twice the typical photo-$z$ uncertainty. In the case of DES \redmagic\ galaxies, the LOS slicing was found to be appropriate for slices of thickness $2s_{v}\approx100~h^{-1}{\rm Mpc}$ for photo-$z$ errors at the level of $\sigma_{z}/(1+z) \approx 0.02$ or $\sim50~h^{-1}{\rm Mpc}$ at $z\approx0.5$. The determination of void parameters then includes a process of circle-growing around void center candidates, and assignment of void radii where the mean density is reached. The last step in the production of void catalogues includes a pruning which is designed to remove multiple detections of a single physical underdensity in neighboring slices. For further details, see \cite{Sanchez2016} who empirically found that $\sim50\%$ of the voids are subject to multiple detections.

We then inverted the void finder algorithm by \cite{Sanchez2016} to find {\it superclusters}. We adopted the smoothed density field that we used for void finding, but this time selected the most over-dense pixels as supercluster center candidates, and grew circles around them until the mean density is reached. This is a rather crude and simplified definition and technique because superclusters typically have non-spherical shape often with multi-spider morphology \citep[e.g.]{Einasto2011}, but for completeness we analyzed the resulting catalogues.

We created super-structure catalogues using shifted ``slicings" of the galaxy catalogue for both data and simulations, as explained in \cite{Sanchez2016}. We then tested for consistency among the different resulting catalogues in terms of general catalogue properties and measurement characteristics.

A free parameter in our method is the scale of the initial smoothing applied to the galaxy density field. \cite{Sanchez2016} used $\sigma=10~h^{-1}{\rm Mpc}$ for their void lensing measurement without testing this parameter value in their analysis. We optimized this choice for an ISW measurement using simulations, given the stacked imprint of DES-like catalogues based on different smoothing levels.

\subsection{The Jubilee simulation}

We analyzed data from the Jubilee ISW project \citep{Watson2014} to estimate the $\Lambda$CDM expectation for the stacked ISW signal of super-structures, following \cite{Hotchkiss2015}. The Jubilee ISW project is built upon the Jubilee simulation, a $\Lambda$CDM (WMAP-5 cosmology) N-body simulation with $6000^3$ particles in a volume of ($6h^{-1}$ Gpc)$^{3}$. We note that the abundance of voids does depend on the cosmological model but given the expected uncertainties in the corresponding ISW signals the difference between WMAP5 and {\it Planck} cosmologies is not important.

The Jubilee simulation is ideal for analyzing the ISW effect because of its large size and relatively high resolution. Specifically, the large box size allows a light cone to be constructed that requires no tiling of the simulation box out to a redshift of $z = 1.4$. Therefore, full sky maps of the temperature anisotropies induced by the ISW effect can be constructed that will not suffer from a cutoff of power on the largest angular scales. Such modes could seriously affect the ISW analyses of existing DES mock catalogues. 

The Jubilee maps of the ISW-induced temperature anisotropies were constructed using a semi-linear approach \citep{CaiEtAl2014} by propagating light rays through the simulation box and obtaining the sky maps of the temperature shift along different directions as seen by a centrally located observer. These maps were pixelized using the \texttt{HEALPix} package at resolution $N_{\rm side}= 512$.

A full modeling of the stacking analysis with Jubilee requires realistic mock galaxy catalogues similar to those in which real voids and superclusters are identified. The individual particle masses of $7.5\times 10^{10} M_{\odot}$ and a minimum resolved halo mass (with $\approx20$ particles) of $\approx 1.5 \times 10^{12} h^{-1} M_{\odot}$ is suitable to perform halo occupation distribution (HOD) modeling of LRG tracers, as discussed by \cite{Watson2014}. The redshifts of the LRGs we considered include Doppler terms, and we also modeled the effect of photo-$z$ uncertainties.

This LRG mock was first designed to model the properties of SDSS LRGs studied in \cite{Eisenstein2005}, and then \cite{Hotchkiss2015} modeled SDSS DR7 LRG data and mocks by \cite{Kazin2010} with a subset of the Jubilee LRGs.

While both the Jubilee LRG mock and the DES \redmagic\ galaxy catalogues are approximately volume limited, there are differences in the number density. The Jubilee mock provides a sample with $\bar{n}\approx8\times10^{-5}h^{3}$ Mpc$^{-3}$ that is lower than the corresponding \redmagic\ values. We chose the high luminosity data for our measurement with $\bar{n}\approx2\times10^{-4}h^{3}$ Mpc$^{-3}$ because it offers more realistic modeling using Jubilee. As a further advantage, the high luminosity sample also traces a larger volume with a fairly homogeneous sampling compared to its high density alternative.

The lower number density of galaxies in Jubilee means that this simulation does not precisely model the DES \redmagic\ population, which could affect our conclusions about the optimal stacking strategy. In sparser galaxy tracers, the number of voids identified decreases, but the average reported void size increases \citep{Sutter2014_DM,Nadathur2015}. More importantly, voids resolved by sparse galaxy samples also on average trace shallower but larger dark matter underdensities \citep{Nadathur2015}, which should have a longer photon travel time and therefore correspond to larger ISW temperature shifts. This conclusion is corroborated by the findings of \cite{Hotchkiss2015}, who examined the ISW effects for voids in two mock LRG catalogues with differing brightness and sparsity in the Jubilee simulation, and found that the sparser sample gave consistently larger $|\Delta T|$. They also found a similar effect for superclusters. We concluded that the expected stacked ISW signal we determine from Jubilee will be an \emph{overestimate} of that observable from superstructures in the DES \redmagic\ data. However, for the given galaxy number densities the difference in expected ISW signals is expected to be relatively small and certainly below the level of noise in the measurement.

\begin{figure*}
\begin{center}
\includegraphics[width=175mm]{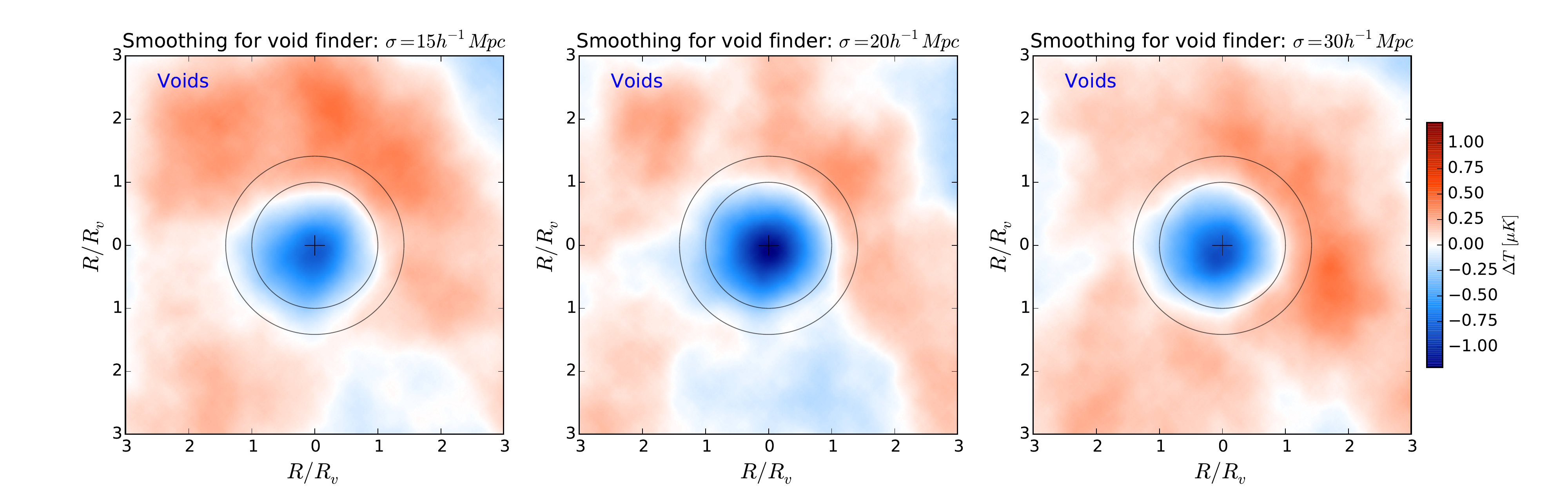}
\label{stacked_images_CV}
\caption{Stacked ISW imprint of full-sky mock Jubilee voids as a function of the initial smoothing. We used $2133$ voids which is the total number of objects for $\sigma=30~h^{-1}{\rm Mpc}$ smoothing. We ordered the voids in the other catalogues by void radius, and considered only the largest $2133$ objects in the stacking for this comparison.}
\end{center}
\end{figure*}

\begin{figure*}
\begin{center}
\includegraphics[width=138mm]{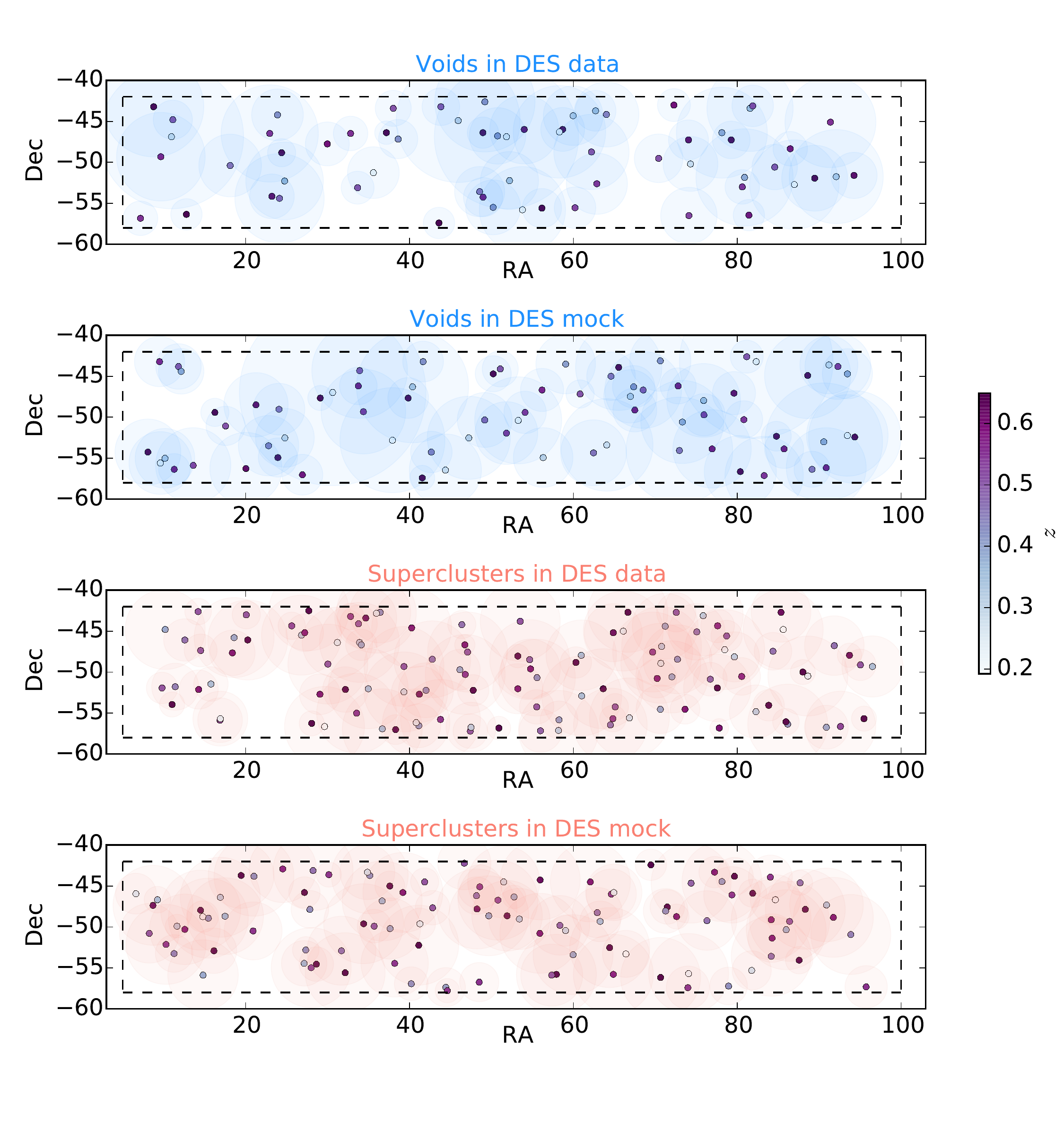}
\caption{Map of the void catalogue (top panels) and the supercluster catalogue (bottom panels) in Y1A1 redMaGiC data and in the Buzzard Y1A1 \redmagic\ mock. We applied $\sigma=20~h^{-1}{\rm Mpc}$ initial Gaussian smoothing to the density field as discussed in the main text. The actual area used for the analysis is marked by the dashed rectangles. Colored disks mark the full angular size of the objects, while colored points in the disk centers indicate the redshift assigned to each void's center. The intensity bar shows the redshifts.}
\end{center}
\end{figure*}

\begin{figure*}
\begin{center}
\includegraphics[width=175mm]{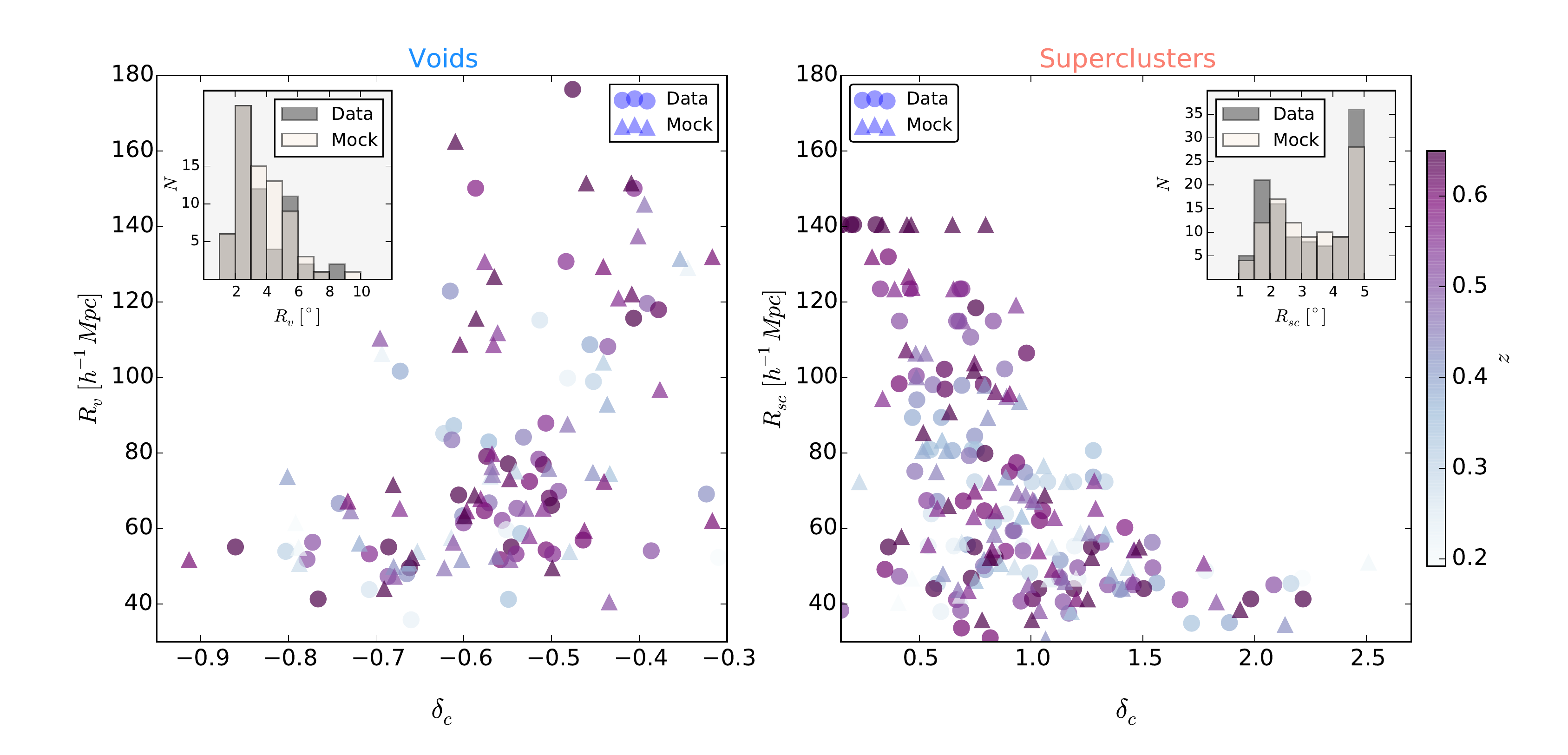}
\label{params}
\caption{Summary plot of parameters: transverse radius $R_{v/sc}$, central under-density $\delta_{c}$, and central redshift $z$ is shown for voids in the left panel and for superclusters on the right side using data (circles) and the mock catalogue (triangles). The insets show the angular size distribution of the objects. Note that unlike in the case of \texttt{ZOBOV} objects, the largest voids we defined are not the most underdense ones.}
\end{center}
\end{figure*}

\section{Modeling the ISW imprint of super-structures}

There are a large number of open choices in how the stacking technique is performed, including how voids and superclusters are defined, catalogue pruning, aggressiveness of masks, and methodology details such as filter size and the total number of objects considered for drawing conclusions \citep[see e.g.][]{Hotchkiss2015}. Prior to looking at the DES data, we first used simulations to minimize the effects of the posterior selection of such parameter values without formally carrying out a blinded analysis. We optimized the signal-to-noise of the ISW measurements by varying the exact methodology of the void finder phase and the stacking procedure.

\subsection{Optimizing the initial smoothing scale}

We tested different values for the initial Gaussian smoothing of the galaxy density field to define void and supercluster centers. We expect that the best possible number is larger than the $\sigma=10~h^{-1}{\rm Mpc}$ value considered by \cite{Sanchez2016}. The ISW detection is sensitive to tracing the full extent of large underdensitites, and larger smoothings automatically merge smaller sub-voids into larger voids, albeit with some uncertainty in the centering and size estimates simply due to the void finding algorithm and limitations of the data \citep{Sanchez2016}. Also, a large smoothing removes smaller void candidates often residing in over-dense environments \citep{CaiEtAl2014}, which however are not expected to significantly contribute to the ISW signal. On the other hand, too coarse smoothing can increase the uncertainties in the position and size estimates because in reality (super)voids are not always spherical and some information about their sub-structure might be informative. 

To optimize the smoothing, we defined void catalogues for the full-sky Jubilee LRG mock catalogue by considering $\sigma=15~h^{-1}{\rm Mpc}$, $\sigma=20~h^{-1}{\rm Mpc}$, and $\sigma=30~h^{-1}{\rm Mpc}$ initial smoothings, and created stacked images of the mean ISW imprint of the structures, as shown in Figure 2. We added Gaussian photometric redshift noise with $\sigma_{z}/(1+z) \approx 0.02$ to the Jubilee redshift coordinates in order to model the \redmagic\ photo-$z$ properties. Additionally, we applied the $0.2 < z < 0.65$ \redmagic\ survey window cut to the Jubilee LRGs to better represent the observational conditions.

We also removed super-structures that exceed the size of the objects that could be detected in the DES data. The full-sky analysis of the Jubilee mock catalogue allowed the finder to identify more extended structures that are practically undetectable with a rather narrow DES Y1-like survey footprint. We excluded $\sim10\%$ of the voids in all simulated cases. In section 4, we will further analyze the importance of the DES Y1-like survey footprint in terms of cosmic variance.

In the example shown in Figure 2, we compared the ISW signals of mock voids and found that the choice of $\sigma=20~h^{-1}{\rm Mpc}$ provides the best contrast and highest absolute value for an ISW imprint. We note that this is an estimate of the signal, not the signal-to-noise ratio. In principle, a more densely populated catalogue of super-structures might be more efficient in reducing the CMB noise. However, we have found that the low absolute ISW signal detected using $\sigma=15~h^{-1}{\rm Mpc}$ is not balanced by the reduced noise levels, and the $\sigma=20~h^{-1}{\rm Mpc}$ choice gives $\sim30\%$ higher $S/N$.

We then applied these findings to the DES Y1 data and mock catalogues. We smoothed the sliced DES Y1 galaxy density fields with $\sigma=20~h^{-1}{\rm Mpc}$ in data and in the mock, and found that the total number of voids is $52 <N_{\rm v}<61$ for 5 different slicings using shifted $z$-bin edges, while supercluster detections are in the range $102 <N_{\rm sc}<111$.

Sizes, locations, and a comparison of important super-structure parameters transverse radius $R_{v/sc}$, central under-density $\delta_{c}$, and central redshift $z$ for data and the mock catalogue are presented in Figures 3 and 4. See descriptions in \cite{Sanchez2016} for further details about the estimation of these void and supercluster properties.

\begin{figure*}
\label{elong}
\includegraphics[width=160mm]{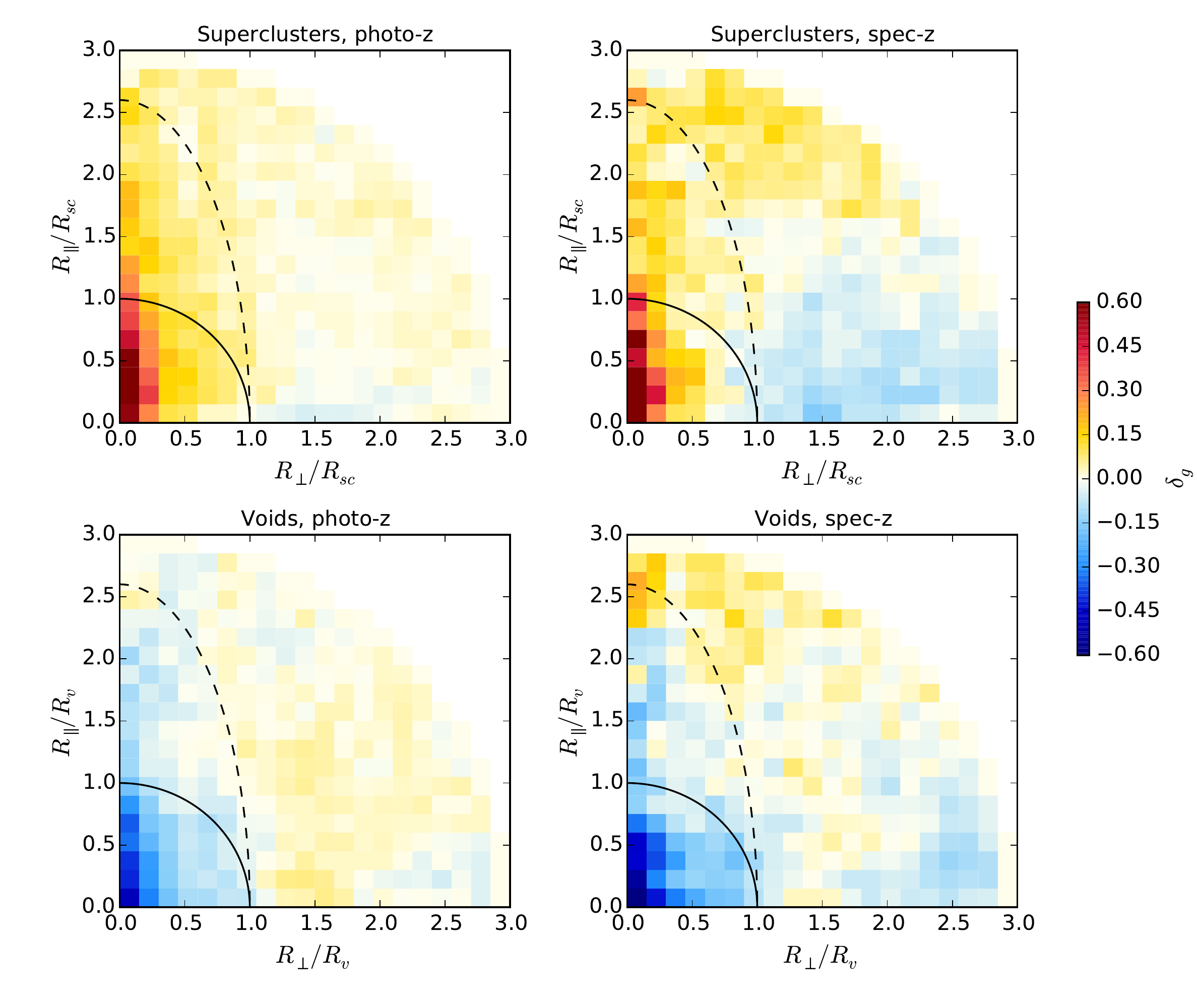}
\caption{The right panels show the {\it true} galaxy density distributions about the selected locations (bottom-right: voids, top-right: superclusters), whereas the left panels show the {\it apparent} distributions when the true galaxy positions are distorted by photometric redshift errors (bottom-left: voids, top-left: superclusters). We performed this analysis using the Buzzard simulation of the Y1A1 \redmagic\ mock catalogue. Solid circles correspond to a spherical super-structure shape while dashed ellipses mark an elongated model with $R_{\parallel}=2.6 R_{\perp}$ estimated for the Gr08 supervoids.}
\end{figure*}

\subsection{Analyses of line-of-sight elongation}

Super-structures elongated in our line-of-sight have a longer photon travel time compared to the spherical case and therefore correspond to larger ISW temperature shifts \citep{MarcosCaballero2015}. It is crucial to understand any biases in the void identification in order to correctly measure and interpret localized ISW localized ISW imprints of large and elongated underdensities. However, it is worth noting that \cite{Flender2013} concluded that the assumption of sphericity does not lead to a significant underestimate of the ISW signal.

In principle, even spherical voids can appear elongated in the line-of-sight in the presence of any photo-$z$ uncertainty for the tracer galaxies. The smearing effect of photometric redshift uncertainties can be reduced when considering LRG tracer catalogues with more accurate photometric redshifts, e.g. the DES \redmagic\ sample. Most of the significant voids are expected to be detected but corrections are required to obtain their true shape parameters. In more extreme cases, \cite{BremerEtal2010} showed that a photo-$z$ smearing at the $\sigma_{z}=0.05(1+z)$ level can easily result in non-detections of typical voids in average environments.

On the other hand, void finders run on photo-$z$ data appear to be more sensitive to systems of multiple voids lined up in our line-of-sight, or underdensities elongated in this preferred direction \citep{Granett2015}. In any case, the analysis of DES-like super-structures in Jubilee provides a realistic and accurate estimate of the $\Lambda$CDM expectation for the ISW imprint of these elongated objects.

We reconstructed the mean shapes of DES voids and superclusters using both their {\it photo-$z$ and spec-$z$} coordinates available in the Buzzard \redmagic\ mock catalogue. The analysis of spec-$z$ coordinates reveals the real shape of the objects defined using photo-$z$ data. Super-structures were selected in the simulation with the same criteria as in the data. We introduced our comparisons of transverse and line-of-sight profiles in Figure~5.

\begin{figure*}
\begin{center}
\includegraphics[width=175mm]{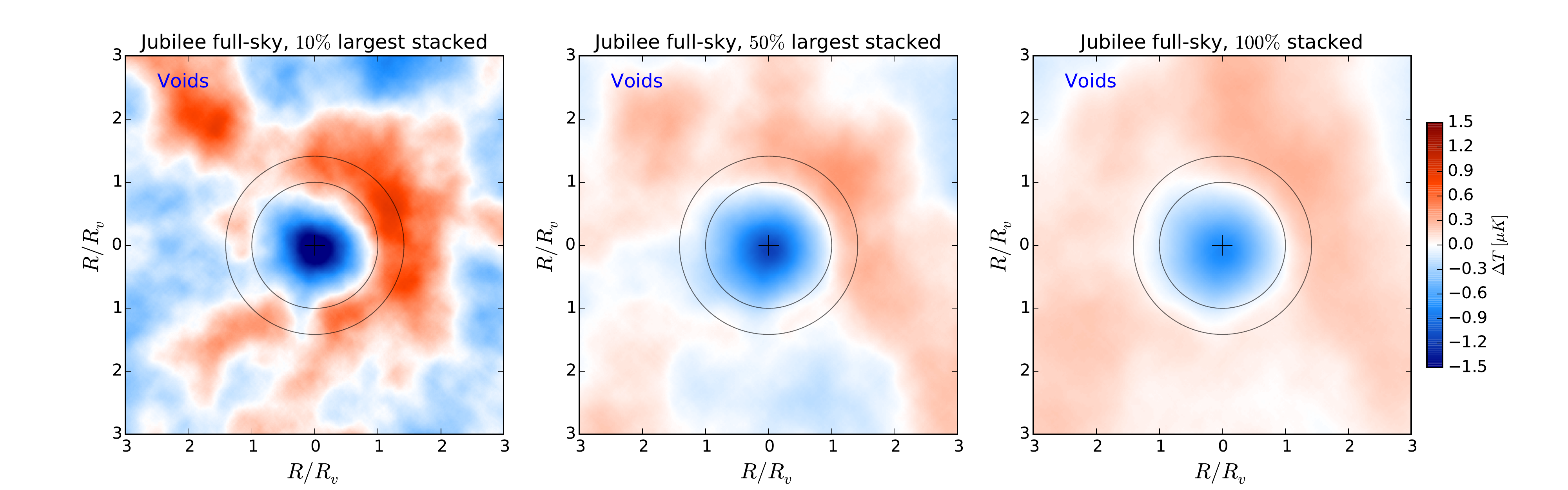}
\includegraphics[width=175mm]{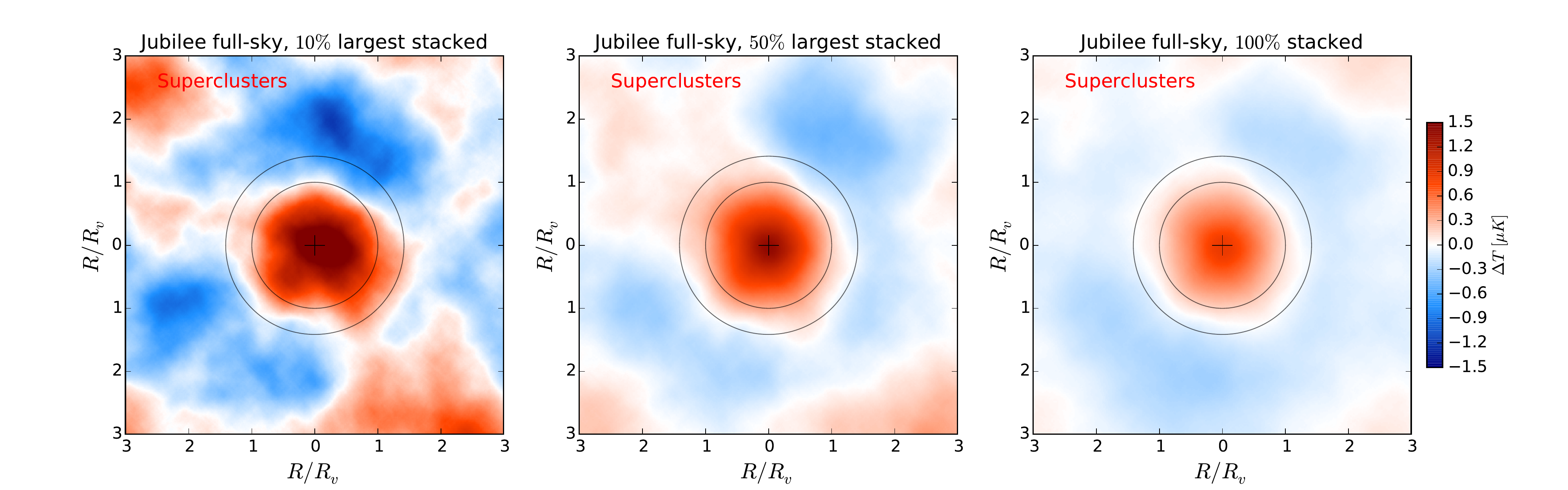}
\label{stacked_images}
\caption{Stacked ISW imprint of Jubilee voids (top) and superclusters (bottom) when selecting of the $10\%$ largest fraction of the voids and superclusters (left), the larger half of the samples (middle), or by stacking all objects (right). Images are shown in units of re-scaled to super-structure radii. Inner circles mark the super-structure radius ($R/R_{v}=1$), while the outer circles mark the boundary of the corresponding CTH filter ($R/R_{v}=\sqrt{2}$). We discuss the observable trends in the imprints in the main text. No smoothing was applied to the Jubilee ISW-only map.}
\end{center}
\end{figure*}

In practice, we measured galaxy densities in $0.15 R_{v}~h^{-1}{\rm Mpc} \times 0.15 R_{v}~h^{-1}{\rm Mpc}$ cells around super-structure centers. We then created a stacked profile in the units of the super-structure radii. We found that the density fields, shown in Figure 5, are inconsistent with the spherical void hypothesis. The measurements, are instead consistent with {\it elongated} objects for both photo-$z$ and spec-$z$ counts. For the supercluster sample, we similarly found that the density map is consistent with structures elongated in the LOS. The similarity of super-structure catalogue properties for data and for the mock catalogue allows us to conclude that our super-structures in observational data are also elongated in the LOS.

In both cases, we found that the elongation is more pronounced when using photo-$z$ coordinates (especially for voids). We note that the elongation is partially the consequence of the void finder algorithm because we consider cylindrical structures by definition. However, \cite{Granett2015} showed that \texttt{ZOBOV} voids also show elongation in the presence of photo-$z$ errors thus we assume that the main cause is the latter.

Analyses of the more informative spec-$z$ coordinates revealed an approximate mean elongation $R_{\parallel}/R_{\perp} \approx 2.2$ for voids and $R_{\parallel}/R_{\perp} \approx 2.6$ for superclusters. We note that the stacked supercluster density profile becomes more compact when using spec-$z$ coordinates. Voids, however, are very similar in angular size using either photo-$z$ or spec-$z$ coordinates but they appear to be $\sim 10\%$ less elongated when spec-$z$ coordinates are used.

We note that the shape analysis of Gr08 super-structures by \cite{Granett2015} revealed qualitatively similar properties with $R_{\parallel}/R_{\perp} \approx 2.6\pm 0.4$ mean elongation in the LOS. In contrast, for voids in the BOSS spectroscopic data, \cite{Nadathur2016} found smaller average ellipticities, and with a random orientation of void major axes relative to the line of sight.

The level of the bias towards elongated objects might depend on the value of the photometric redshift uncertainties, on the initial smoothing applied to the density field, and/or on the void finder algorithm itself. Gr08 used an SDSS photo-$z$ catalogue with $\sigma_{z}/(1+z) \approx 0.05$ uncertainties while in our case the photo-$z$ scatter for the \redmagic\ sample is $\sigma_{z}/(1+z) \approx 0.02$. This better line-of-sight resolution of DES data implies sensitivity to objects with less elongated shape that are otherwise smeared out by larger photo-$z$ errors.

For completeness, we mention galaxy ``troughs" that represent the most extreme case in this comparison of elongated underdensitites. They are defined as the most underdense regions of thick projected density fields, for instance 0.2<z<0.5 in \cite{Gruen2016}. By construction, the minimal use of photo-$z$ information results in detections of underdensities biased towards very elongated LOS shapes or superpositions of various underdensities along the LOS.

\begin{figure*}
\begin{center}
\includegraphics[width=175mm]{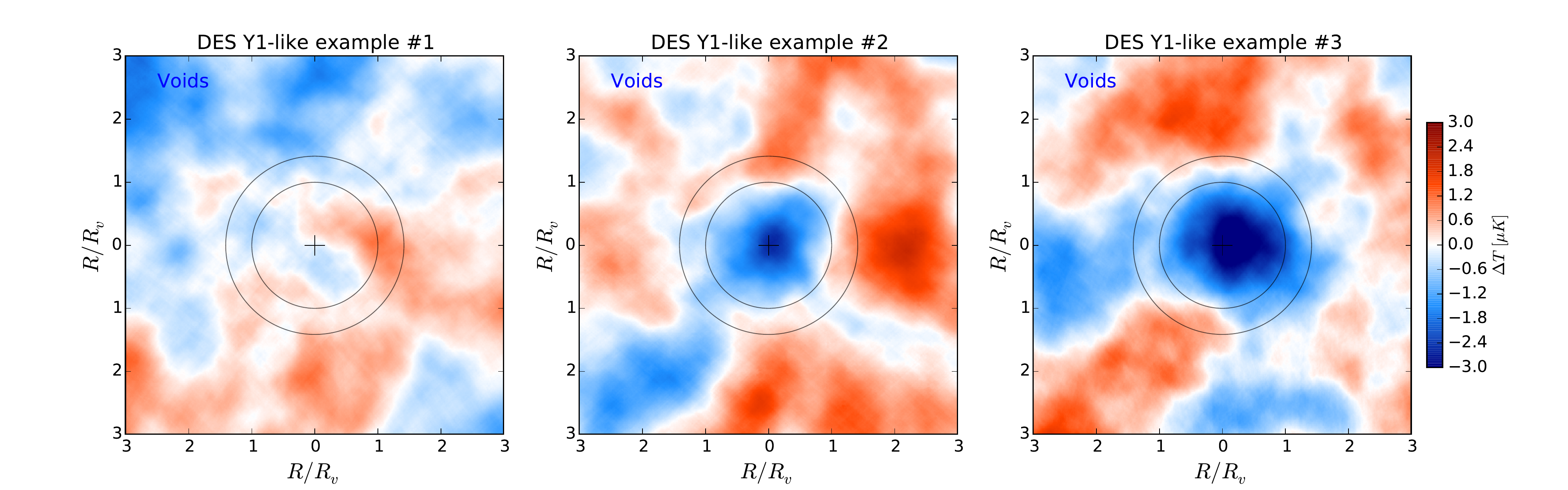}
\label{stacked_images_CV}
\caption{Examples of stacked ISW imprints of a posteriori selected {\it DES Y1-like} void catalogues in Jubilee. We stacked $100\%$ of the voids in this example; thus one can compare these results to the top-right panel of Figure 6 that shows the full-sky estimate. Note the high variability of the shape and amplitude of the signal when measured at different parts of the sky. Inner circles mark the super-structure radius ($R/R_{v}=1$), while the outer circles mark the boundary of the corresponding CTH filter ($R/R_{v}=\sqrt{2}$). }
\end{center}
\end{figure*}

\subsection{Most significant super-structures}

The ISW signal expected in $\Lambda$CDM is so small that it is dominated by the primary anisotropies even with stacking applied to these individually noisy measurements. However, the voids and superclusters identified in the DES footprint are excellent candidates for a follow-up analysis using independent data,  even if the expectation for the signal-to-noise appears to be low.

In principle, since the expected ISW imprint is smaller for the smallest objects, it is possible that using all of these voids is not optimal for an ISW detection. Furthermore, smaller voids have the highest noise level in the sample because they come with small filter size where the filtered CMB variance is larger, even if the intrinsic CMB variance is actually smaller at smaller scales. Differential binning of the super-structures and special weighting techniques based on inverse variance or signal-to-noise are possible, but in our case such corrections are difficult because of the small sample size. Instead, we advanced the existing stacking and pruning methodologies \citep{CaiEtAl2014,Nadathur2015} by adding a $S/N$ analysis to our measurement pipeline to {\it a priori} decide, based on simulations, what is the most useful part of the data to include.

While environmental effects, differences in density profiles and redshifts, and different shapes can be important for reliable estimates of the ISW imprint of super-structures, the signal is expected correlate with the radius. We therefore ordered the Jubilee voids and superclusters in our full-sky catalogues by their radii, and cumulatively measured the stacked imprint of their subsets in $10\%$ bins.

We first measured the ISW imprint of mock super-structures, as shown in Figure 6. We re-scaled the images based on the angular size of the objects in order to test the total extent of the ISW imprints compared to the angular size of the super-structures. We compared the imprint of the $10\%$ largest fraction of the voids and superclusters to the imprint of the larger $50\%$ of the samples, and then to the imprint detected by stacking all objects. We observed that the absolute values are similar for voids and superclusters in all cases. However, as expected, there is a clear trend of more significant imprints with larger $|\Delta T|$ for larger objects. The imprint of the largest superstructures, on the other hand, is more noisy due to the low number of objects in the stacking, even considering the ISW-only map without CMB noise. 

These findings are comparable to the results by \cite{Hotchkiss2015} who analyzed two different void populations using the Jubilee mock and the ISW map.

\subsection{Cosmic variance and large-scale modes}

The previous estimates that we have obtained are based on analyses of {\it full-sky} Jubilee mock catalogues. While this approach is quite helpful to find the correct $\Lambda$CDM expectations, the small size of the DES Y1 survey area is important because the ISW imprints, even when stacked, can vary significantly in small patches across the sky. Therefore, we measured the stacked ISW imprints of $100$ randomly placed DES Y1-like patches in order to test the fluctuations of the signal. Note that these patches are not totally uncorrelated because only $\approx40$ independent DES Y1-like patches could be placed on a full sky map.

\begin{figure}
\begin{center}
\includegraphics[width=85mm]{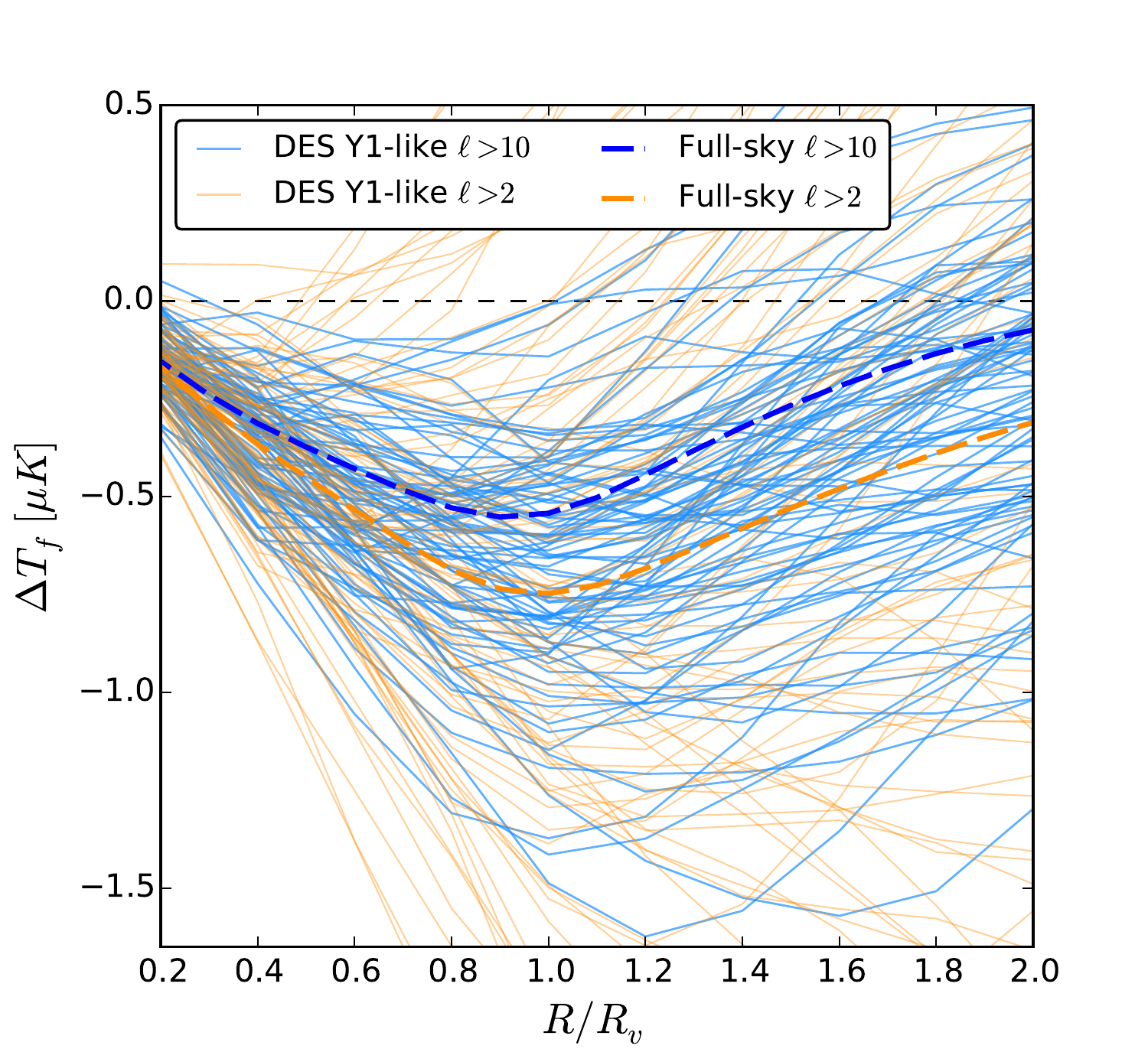}
\label{profs_CV}
\caption{Stacked measurements of ISW signals are compared in 100 DES Y1-like Jubilee patches (thin solid curves) to the full sky Jubilee estimate (single thick dashed curve). The curves show the individual filtered ISW imprint as a function of the filter size for the small DES Y1-like patches and for the full-sky estimate. The higher variability of the orange curves (with $\ell>2$ modes in the ISW map instead of only $\ell>10$) highlights the effect of the removal of large-scale modes from the ISW map to reduce the field-to-field fluctuations.}
\end{center}
\end{figure}

\begin{figure*}
\begin{center}
\includegraphics[width=88mm]{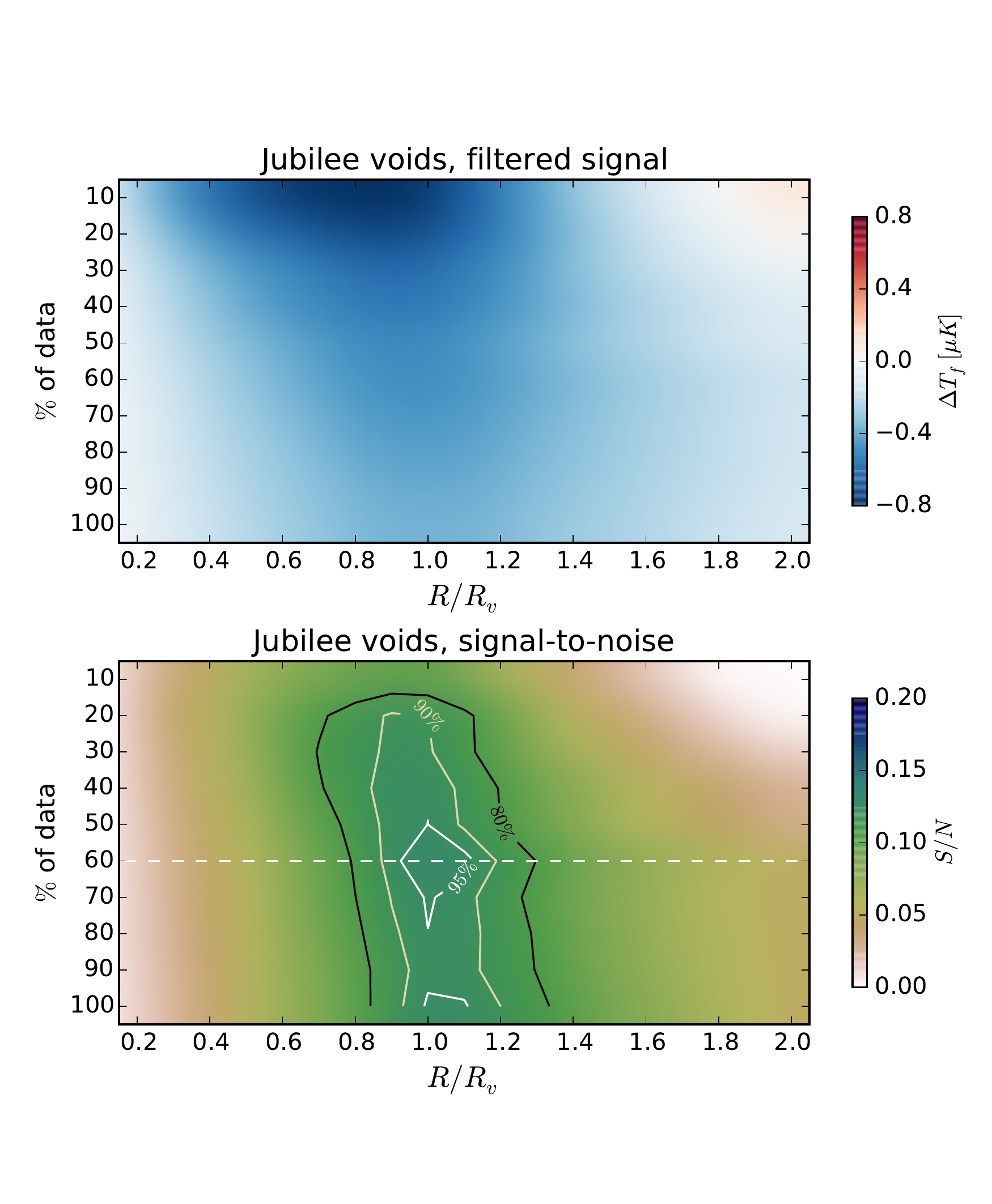}
\includegraphics[width=88mm]{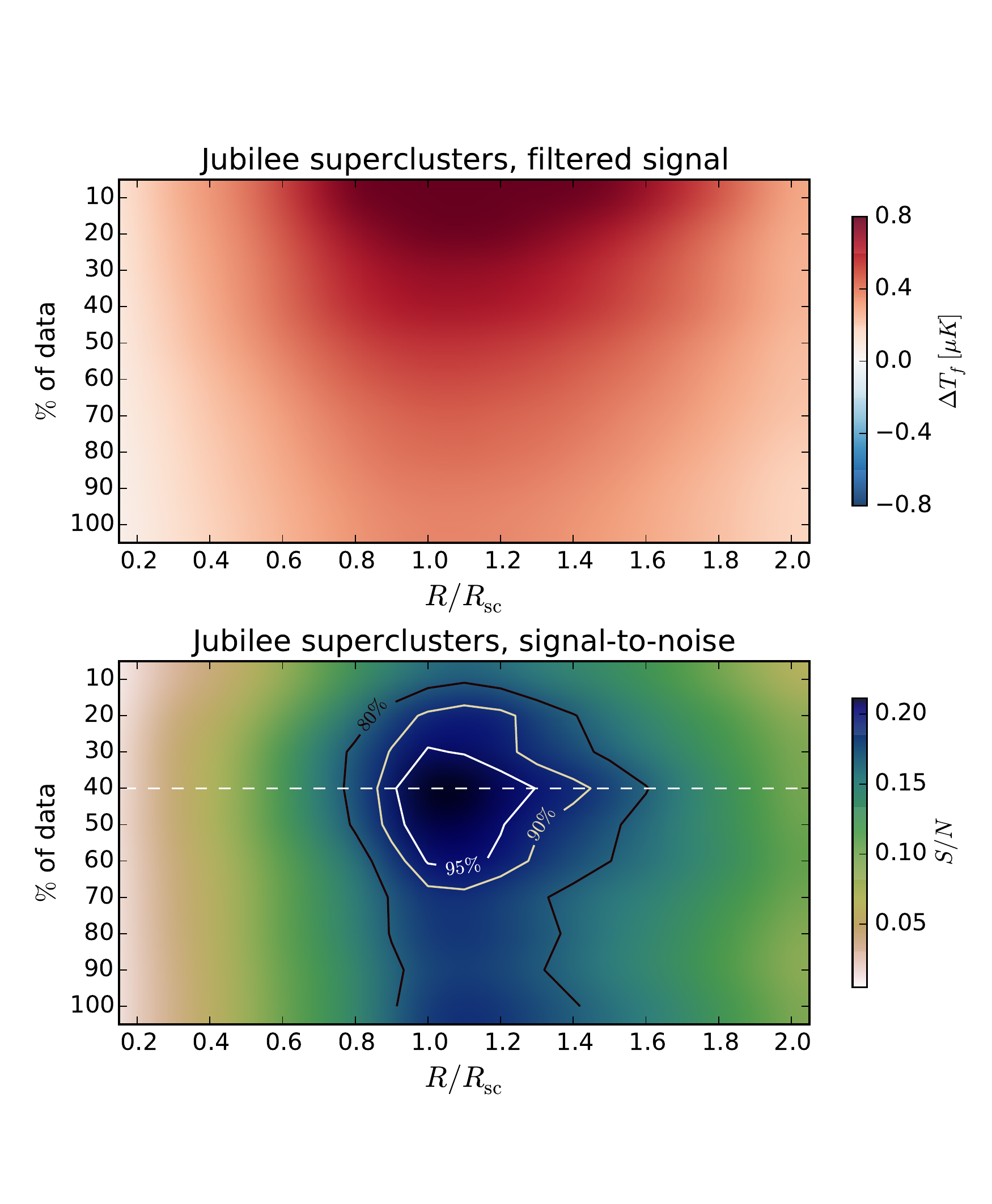}
\label{mainres}
\caption{Top panels show the stacked ISW-only signals using data fraction and filter size indicated by the axes. We considered the {\it full-sky} Jubilee signals for this test. We estimated the noise levels by considering the angular sizes and positions of $N_{\rm v}=52$ voids and $N_{\rm sc}=102$ superclusters that are detectable in a DES Y1-like volume with our methods. The bottom panels show the signal-to-noise ratios given the properties of real-world DES Y1 super-structures. The contours mark pixels with $95\%$, $90\%$, and $80\%$ of the $S/N$ maxima. At $60\%$ for voids and $40\%$ for superclusters, the dashed lines indicate the data fraction with highest (but still very low) $S/N$.}
\end{center}
\end{figure*}

Another important ingredient in this analysis is the role of the large-scale modes in the ISW map. These long wavelength fluctuations can bias and distort the ISW measurements in relatively small survey windows, thus some previous studies have already considered the effects of their removal \citep{Hotchkiss2015, CaiEtAl2014}. Another motivation to study these modes is the reduced CMB noise level in real data without e.g. the $2<\ell<10$ modes with essentially unchanged signal through a CTH filter \citep{Ilic2013}. 

In Figure 7, we show three examples of the highly variable ISW imprints measured in these relatively small $\sim1000~deg^{2}$ areas in Jubilee, considering only $\ell>10$ modes. In some cases, hardly any ISW imprint is detectable, but in other cases the central imprints reach the $\Delta T\approx-3~\mu K$ level. These are $\sim100\%$ fluctuations compared to the full-sky result with all voids included in the stacking (top-right panel of Figure 6).

We then quantified the variability of the stacked ISW signal using 100 randomly placed DES Y1-like patches in Jubilee instead of only three dissimilar and extreme examples. In Figure 8, we compare the stacked and CTH-filtered ISW signals for Jubilee voids with and without $2<\ell<10$ modes in the ISW map considering different filter sizes in the units of the void radius. We found that $\sim20\%$ of the accessible full-sky ISW signal is lost if these modes are removed. However, the filtered signals show significantly less variation around the full-sky estimate without $2<\ell<10$ modes, thus we concluded that it is reasonable to remove these large-scale modes for DES Y1-like patches for a well-controlled measurement. Note that real-world measurements with peculiar shapes for the filtered signal or high ISW-like amplitude should be compared to extreme cases in this distribution. Supported by these findings, we only considered $\ell>10$ modes in our stacking measurements later in the paper.

\begin{figure*}
\begin{center}
\includegraphics[width=88mm]{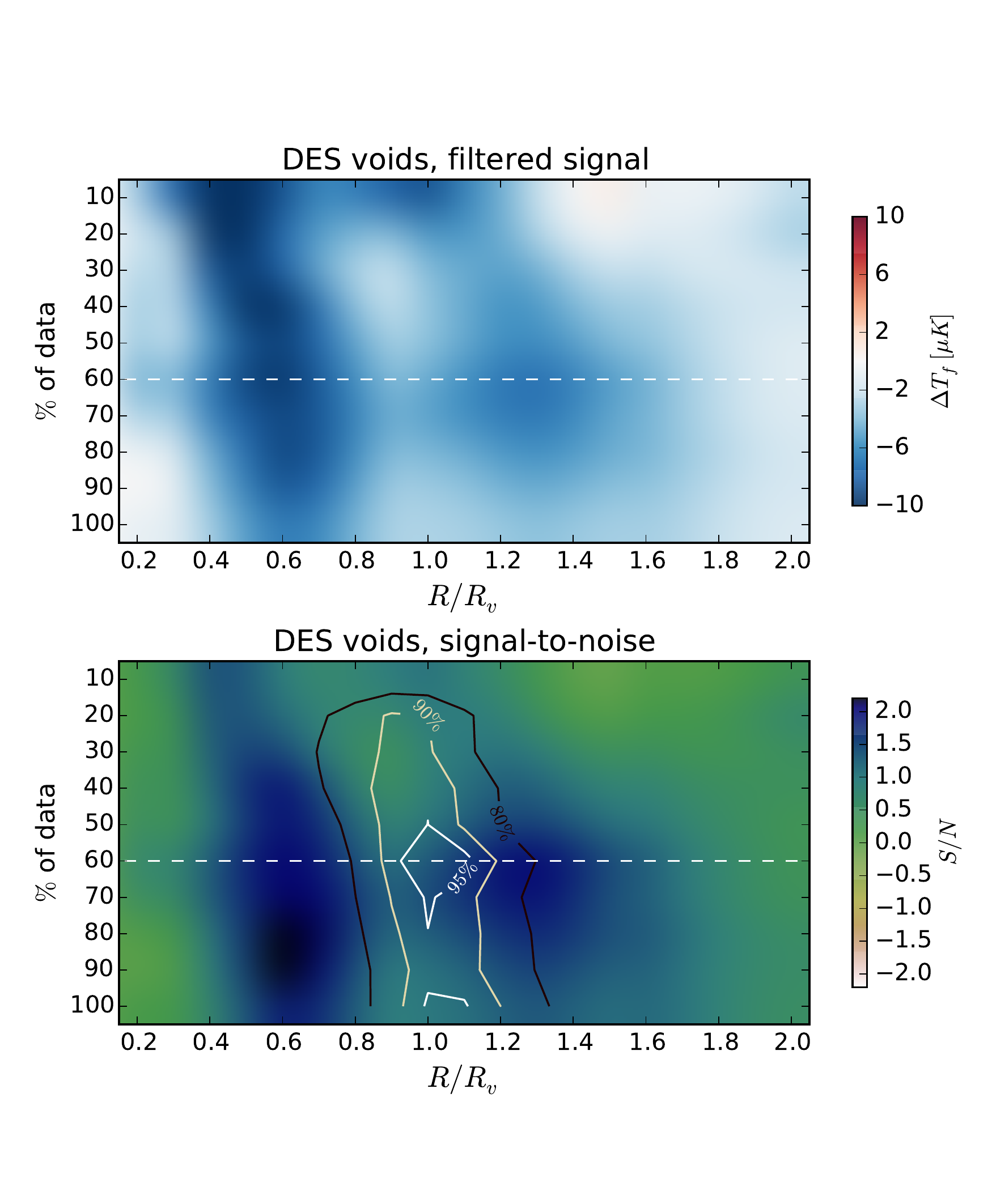}
\includegraphics[width=88mm]{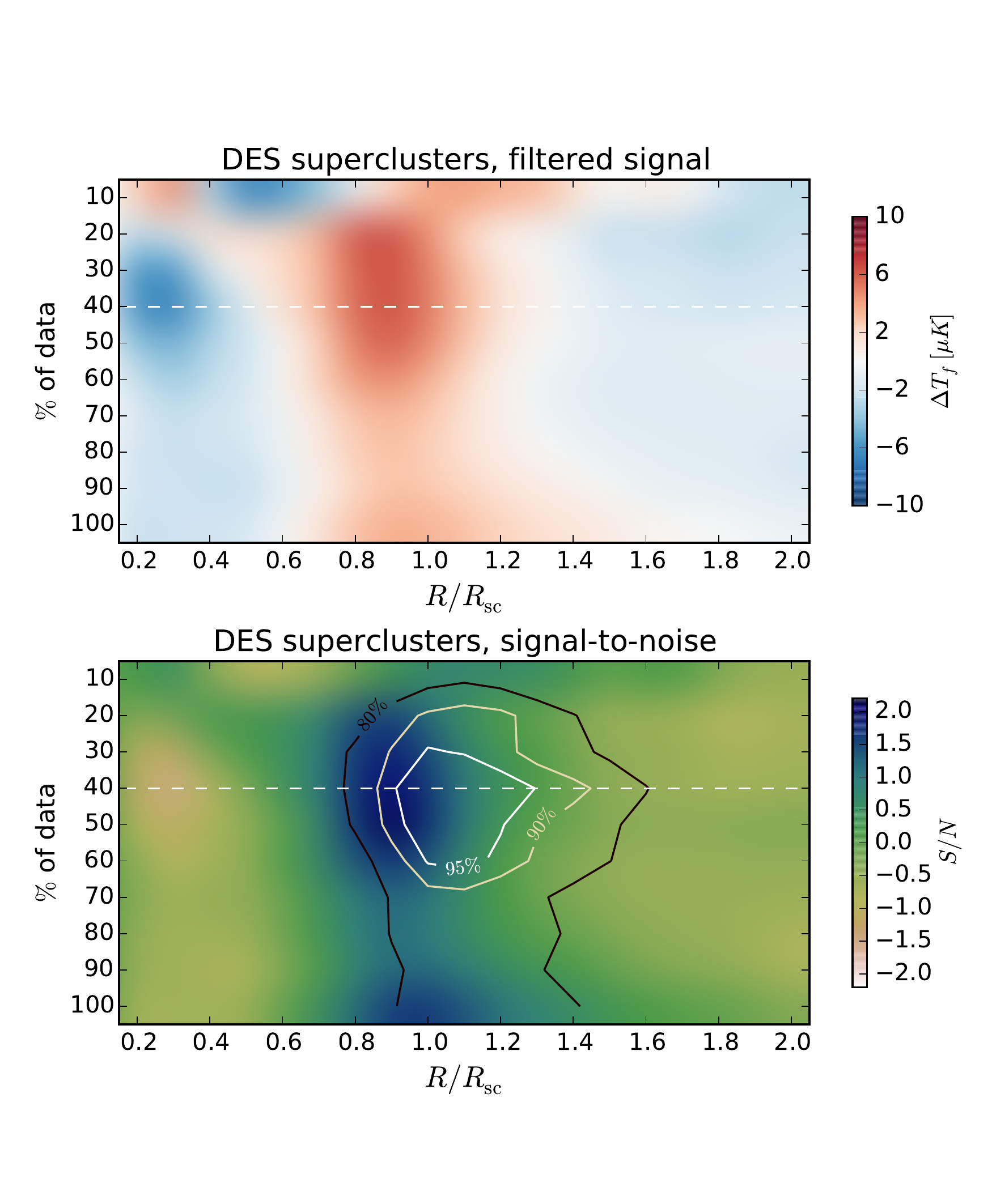}
\label{mainres2}
\caption{Top panels show stacked {\it Planck} CMB signals considering different data fractions and filter sizes, as for Jubilee in figure 10. The top panels shows the stacked CMB imprint of DES Y1 voids and superclusters, while the bottom panels show the signal-to-noise ratios given the noise properties of the super-structures. We over plot contours obtained in the Jubilee analysis in order to show where the maximum $S/N$ is expected. The dashed lines indicate the data fraction with highest $S/N$ based on Jubilee, i.e. the one that we should {\it a priori} consider.}
\end{center}
\end{figure*}

\subsection{Importance of filter size}

In what follows, we closely relied on the methods by \cite{Hotchkiss2015}. We tested for variations in the filtered ISW signal using different filter sizes and by stacking different fractions of the radius-ordered super-structure catalogues. We introduced our findings in Figure 9. Consistently with the findings by \cite{Hotchkiss2015}, we found $|\Delta T_{f}| \leq 1~\mu K$ imprints both for voids and superclusters. The imprints are quite symmetrical, except perhaps the difference in the location of the strongest ISW signals that is observed at $R/R_{v}\approx0.8$ for the largest voids and at $R/R_{v}\approx1.2$ for the largest superclusters. For the whole sample, the best re-scaling factor approaches $R/R_{v}\approx1.0$ both for voids and superclusters. We note that the imprint of Jubilee super-structures is in good agreement with analytical models presented in \cite{Nadathur2012} and \cite{Flender2013}.

We then obtained the signal-to-noise expectation for the ISW-only $|\Delta T_{f}| \leq 1~\mu K$ imprints of given DES Y1 super-structure catalogue properties. The most relevant parameters for this test are the number of voids and superclusters and their angular sizes. We estimated statistical uncertainties by repeating the stacking measurements using $1000$ Gaussian CMB simulations generated with the \texttt{HEALPix} \citep{healpix} \texttt{synfast} routine using the {\it Planck} 2015 data release best fit CMB power spectrum \citep{Planck_15}. Gaussian simulations without considering instrument noise suffice because the CMB error is dominated by cosmic variance on the scales we consider \citep[see][]{Hotchkiss2015}. We decided to follow the strategy of keeping the voids fixed and varying the CMB realization, because in this case the overlap-effects for super-structures are accounted for more efficiently. Potential large-scale CMB features in the DES footprint are not expected to affect our CTH-filtered results at few-degree scales.

The maximal signal-to-noise remains at the $S/N\approx0.2$ level even for the more numerous population of DES Y1 superclusters, as indicated in Figure 9. Such a modest signal is not surprising in the light of the similar findings by \cite{Flender2013} who considered Gr08-like catalogues with variations in the filter radius and in the number of objects. While the low $S/N$ expectations make any detections of ISW(-like) effects unlikely, the anomalously high signals found by Gr08 motivate a follow-up measurement with similar conditions but independent sky coverage.

\section{Stacking measurement with DES super-structures}

In the previous section, we demonstrated the sensitivity of the ISW imprint of Jubilee super-structures to details in the CTH filtering methodology including catalogue construction and the measurement itself. The knowledge of the behavior of the estimated ISW imprints now allows a {\it quantitative} comparison of $\Lambda$CDM predictions and real-world results using the DES data. 

\begin{figure*}
\begin{center}
\includegraphics[width=155mm]{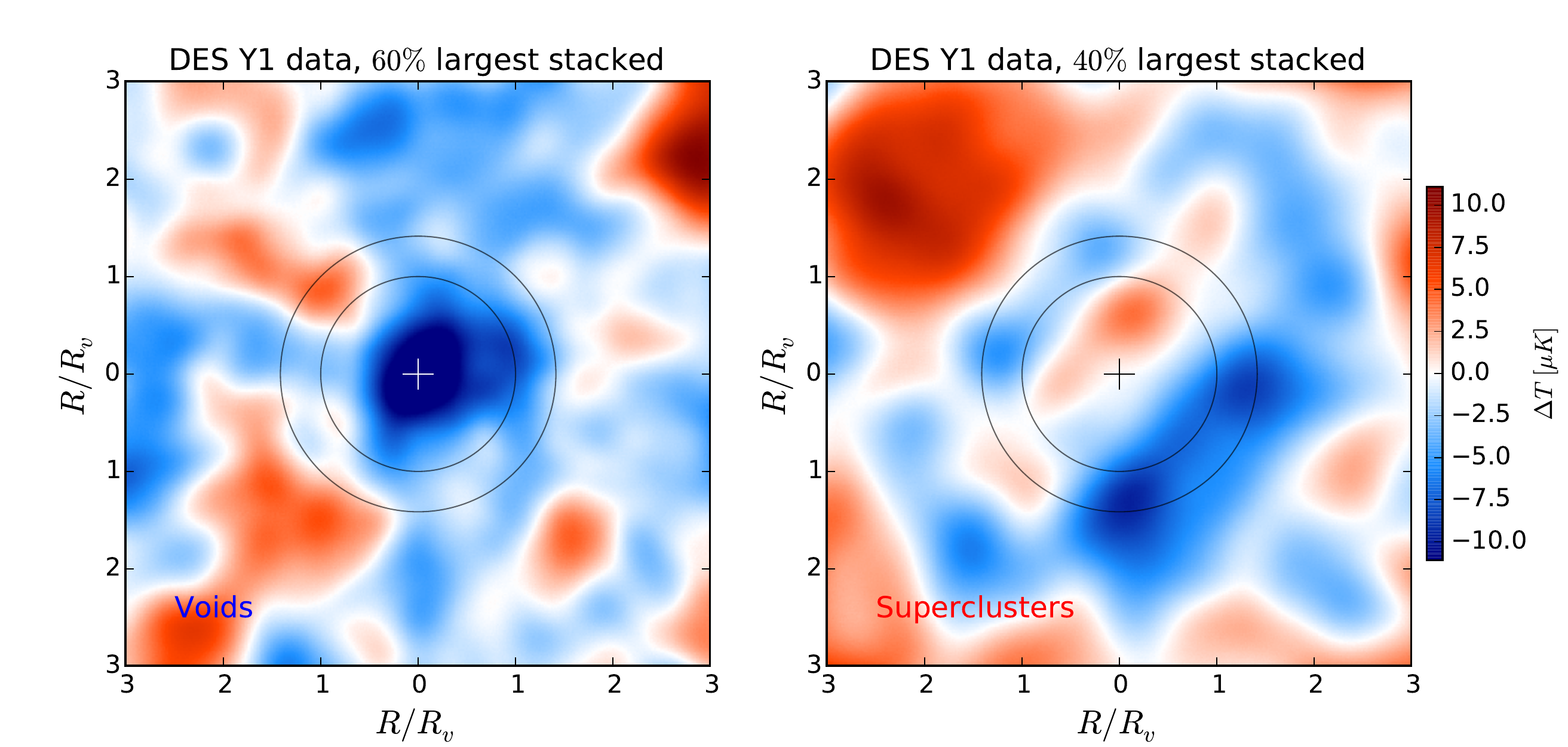}
\label{stacked_images_DES}
\caption{Stacked CMB imprint of DES Y1 voids (left) and superclusters (right). Images are re-scaled to super-structure radii. We applied a smoothing to the individual raw CMB images only for this illustration using $\sigma=3^{\circ}$ symmetrical Gaussian beam in \texttt{HEALPix}. Inner circles mark the super-structure radius ($R/R_{v}=1$), while the outer circles mark the boundary of the corresponding CTH filter ($R/R_{v}=\sqrt{2}$). We discussed the observable trends on the images in the main text. Note the different color scale compared to that of Figure 6.}
\end{center}
\end{figure*}

\subsection{ISW imprint in data vs. simulation}

We used the DES Y1 void and supercluster catalogues introduced in Figures 3 and 4. We followed the same procedures to order the objects. We then performed the CTH filtering measurement in the data fraction vs. filter size parameter space that we investigated using Jubilee mocks.

In Figure 10, we directly compared our Jubilee results to the {\it real-world} DES findings. The first immediate observation is the presence of higher fluctuations in the signals compared to the ISW-only Jubilee stacking results. As in the case of Jubilee, we considered the errors described in Section 3.5. $S/N\approx2$ is observed when $90\%$ of the radius-ordered data is stacked, although our Jubilee calibrations predict that the best chance to detect a signal is to stack $60\%$ of the data and to consider filters $R/R_{v}\approx1.0$.

The combination of DES voids and {\it Planck} data showed $\Delta T_{f} < 0~\mu K$ imprint everywhere in the parameter space for the measurement we have explored. With the optimal configuration we found $\Delta T_{f} \approx -5.0\pm3.7~\mu K$. Moreover, a coherent $\Delta T_{f} \approx - 10~\mu K$ imprint emerged close to $R/R_{v}\approx0.6$ using all data fractions, in particular $\Delta T_{f} \approx -9.8\pm4.7~\mu K$ was observed for the optimal $60\%$ data fraction.

The stacking analysis of DES Y1 superclusters (see the right panel of Figure 10) shows similar features when compared to Jubilee results. The highest signal of $\Delta T_{f} \approx 5.1\pm3.2~\mu K$ is observed for the largest $40\%$ of the sample as predicted in the Jubilee analysis. Furthermore, the location of the observed peak in the $S/N$ is close to the $R/R_{\rm sc}\approx1.0$ Jubilee-based prediction.

Although the $R/R_{v}=0.6$ re-scaling parameter resulted in the most significant imprint, we had no {\it a priori} reason to choose it for our conclusions and posterior choices reduce the significance of anomalous features. However, the magnitude of these posteriori selected imprints is similar to the imprint of super-structures found by \cite{GranettEtAl2008} thus worth further investigation, especially because other void catalogues based on spec-$z$ tracers have not shown such higher-than-expected signals. Gr08 supervoids also show the most anomalous CMB imprint considering $R/R_{v}\approx0.6$\footnote{Initially, a wrong value of $R/R_{v}\approx1.2$ appeared in \cite{Ilic2013} but it later has been corrected to $R/R_{v}\approx0.6$ in the journal paper \cite{Ilic_ERRATUM}.} when a stacking analysis with re-scaling is performed instead of the original constant $R=4^{\circ}$ filtering. \cite{CaiEtAl2014}, \cite{Kovacs2015}, and recently \cite{Cai2016} also reported that re-scalings $R/R_{v}\approx0.6$ or $R/R_{v}\approx0.7$ result in excess signals using SDSS DR7, BOSS DR10, BOSS DR12 void catalogues, respectively. However, \cite{Hotchkiss2015} pointed out that this empirical relation does not necessarily hold for all void definitions and it depends on void parameters; thus the importance of these findings is unclear.

Somewhat similarly, the Eridanus supervoid was found to be significantly elongated in the LOS \citep{KovacsJGB2015} and it appears to be aligned with the CMB Cold Spot. However, the predicted ISW imprint profile disagrees with the observed profile of the Cold Spot \citep{Nadathur2014}.

\subsection{Stacked images for DES data}

We continued our analysis by creating a stacked CMB image of the largest $60\%$ of the DES voids and $40\%$ of the DES superclusters. These data fractions correspond to the dashed lines in Figure 9 and 10 that mark the peak location in the Jubilee $S/N$ map.  Therefore, we were guided to make our main conclusions based on this subset of the data.

Figure 11 illustrates the cold imprint of DES voids and a more modest hot imprint of DES superclusters. We observed some level of compensation around the central regions. For voids, the central cold region is the most significant feature in the image, and its shape and compactness suggests a real feature in the CMB data. 

For superclusters, the rather cold ring-like area around the $R/R_{\rm sc}>1.0$ zone contributes to the $\Delta T_{f} \approx 5~\mu K$ CTH-filtered signal coming from this image because the central temperatures only reach $\Delta T \approx 3~\mu K$. Such ISW features are in fact not unexpected because superclusters are typically surrounded by large underdensitites that leave their own negative ISW imprint (see Figure 6). 

These findings highlight the advantage of using Jubilee for modeling the signals, because analytical models typically only predict the ISW imprint of isolated structures, creating a situation that is clearly unrealistic \citep[e.g.][]{FinelliEtal2014}.

\subsection{Consistency of data and simulations}

We next made a detailed consistency test of CTH-filtered signals as a function of the filter radius in simulation and in real-world data. We show our findings in Figure 12. Separately for voids and superclusters, we compared the DES and Jubilee imprints for data fractions selected {\it a priori} based on the $S/N$ analyses (see also Figures 9 and 11). Figure 12 essentially shows the same information on signal and noise as presented at the image level, but this way the actual {\it consistency} of DES data and $\Lambda$CDM predictions becomes directly visible.

The error bars in Figure 12 represent statistical uncertainties obtained by repeating the stacking measurements using $1000$ Gaussian CMB simulations, as explained in Section 3.5. As an error on the model ISW signal, we also show $1\sigma$ fluctuations in filtered ISW signals of individual DES Y1-like patches in Jubilee. This illustrates the possible effects due to the relatively small DES Y1 survey footprint and the corresponding cosmic variance limitation (see also Figure 8). We concluded that fluctuations due to small sky coverage are significant, but too small to explain the high $\Delta T_{f}$ values found in DES measurements.

We inferred that for the optimized $R/R_{v}\approx1.0$ re-scaling value, the DES measurements are consistent with the imprints of the most extreme DES-like patches in Jubilee at the $\sim1.2\sigma$ level. The imprints of superclusters behave similarly. However, we observed a curious negative signal beyond $R/R_{\rm sc}\approx1.5$ that indicated the possible role of extended underdensities around DES superclusters in the measurable ISW-like imprints.

Therefore, we cannot report a detection of highly significant anomalies in the DES data. However, Gr08 and DES super-structure catalogues both show elongation along the line-of-sight and for both samples the $R/R_{v}\approx0.6$ re-scaling maximizes their ISW-like imprint with a similarly high amplitude. Such connections are worth exploring using larger catalogues specially defined using \redmagic\ (-like) photo-$z$ tracer samples. 

\begin{figure}
\begin{center}
\includegraphics[width=85mm]{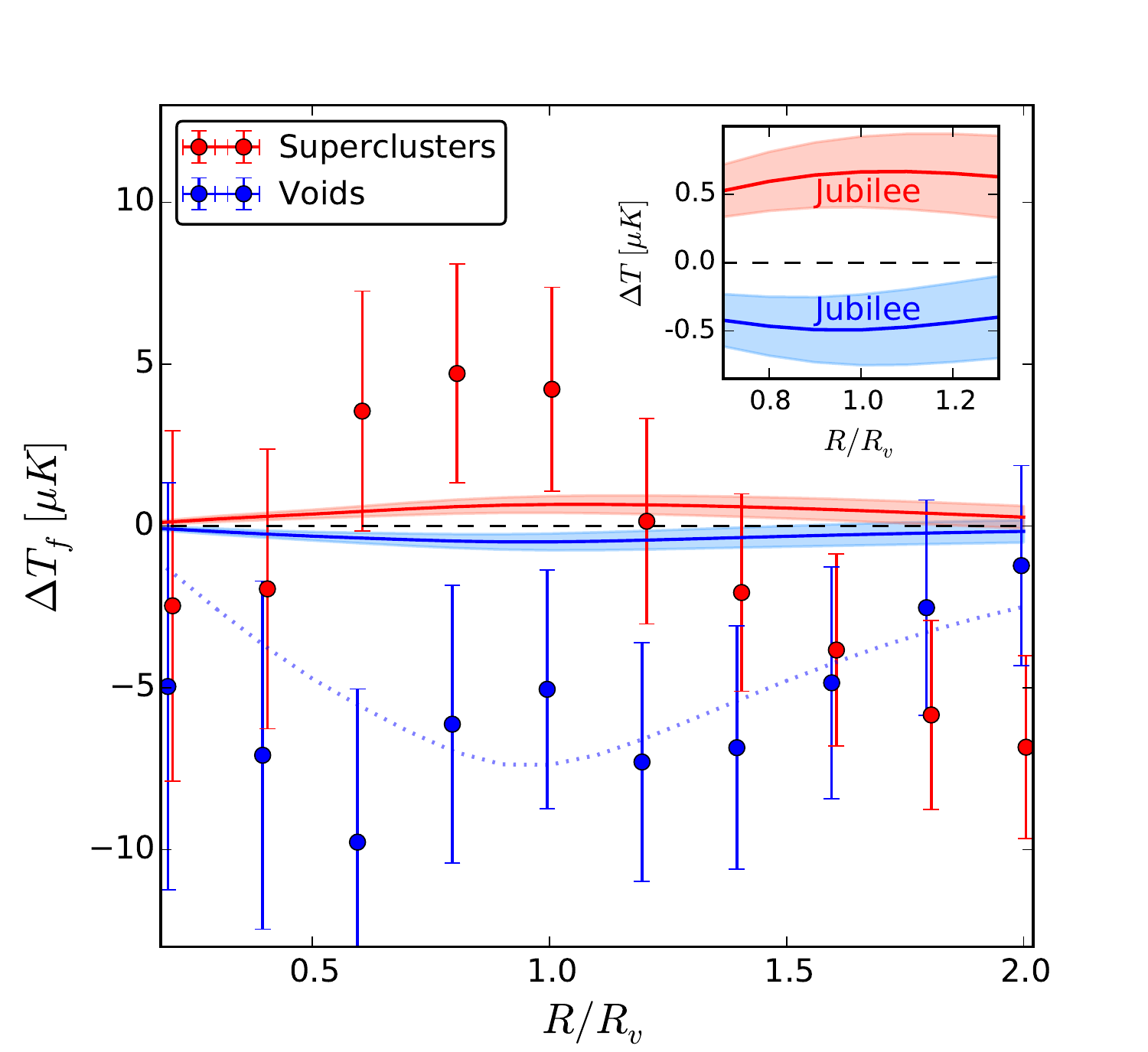}
\label{profiles_comp}
\caption{Filtered signals are compared as a function of filter size for voids (blue) and superclusters (red). Data points and error bars show DES results, while the solid lines are the corresponding Jubilee predictions. The pale blue and pale red shaded regions indicate $1\sigma$ fluctuations around the full-sky estimate in the ISW-only signals if only DES Y1-like patches are considered for the measurement in Jubilee. The inset shows the relation of these models in greater details. The blue dotted line corresponds to an imprint $15\times$ stronger than the actual $\Lambda$CDM prediction for voids based on Jubilee (no fit to data points).}
\end{center}
\end{figure}

\section{Discussion \& Conclusions}

The higher-than-expected ISW(-like) imprints of SDSS super-structures detected by \cite{GranettEtAl2008}, if confirmed, represent a great challenge for standard $\Lambda$CDM cosmology. The evidence for significant LOS elongation of the Gr08 voids also motivates further studies to better understand how void finders perform using photo-$z$ tracer data.

The void finder algorithm developed by \cite{Sanchez2016} represents such an effort, and demonstrates the potential in void science with photometric redshift survey data. The analysis of the DES data provided a great chance to probe the claims by Gr08 with an independent data set. To simplify the interpretation of the measurement, we used the Jubilee simulation to decide, independently of the actual data, which voids and superclusters to stack in our analysis given the catalogue properties and its noise characteristics.

As a $\Lambda$CDM prediction, we found $|\Delta T_{f}| \leq 1~\mu K$ stacked imprint for all data fractions and re-scaled filter sizes for both Jubilee voids and superclusters. This is consistent with previous analyses that estimated the ISW imprint of Gr08-like superstructures \citep[e.g.][]{Flender2013,Hotchkiss2015}. In DES Y1 data, we found $\Delta T_{f} \approx -5.0\pm3.7~\mu K$ for voids using the best configuration motivated by the Jubilee analysis. The most significant $\Delta T_{f} \approx 5.1\pm3.2~\mu K$ stacked imprint for superclusters was detected close to the best filter size and data fraction predicted using Jubilee. The other potentially interesting feature was the $R/R_{v}\approx0.6$ result for DES voids that represents a rather noisy and {\it a posteriori} selected $\Delta T_{f} \approx -9.8\pm4.7~\mu K$ imprint that is $\sim2\sigma$ away from the $\Lambda$CDM predictions.

Expressed in terms of an ISW ``amplitude", we note that our main results are consistent with the $A_{\rm ISW}=1.64\pm0.53$ value (i.e. $\sim1.2\sigma$ higher than the $A_{\rm ISW}=1$ $\Lambda$CDM prediction) reported by \cite{NadathurCrittenden2016}. Our $|\Delta T_{f}|$ results can be turned to a $A_{\rm ISW}\approx8\pm6$ constraint for voids, while for superclusters we found $A_{\rm ISW}\approx8\pm5$. The DES findings are also consistent with the $A_{\rm ISW}\approx6$ detected by \cite{Cai2016} at the modest $\sim1.6\sigma$ significance level.

While these filtered detections have low statistical significance, they do have an amplitude higher than expected in $\Lambda$CDM models. It is interesting to ask whether changes to the cosmological model could explain such differences in the ISW imprint of voids. 
However, \cite{Nadathur2012} concluded that the freedom to vary the $\Lambda$CDM model parameters, given other constraints, is not enough to overcome the discrepancy with observation; \cite{CaiEtAl2014} further found that the Gr08 observation cannot be explained in $f(R)$ models either.

Any excess signal, combined with the anomalous findings of \cite{GranettEtAl2008}, could be more than a chance noise fluctuation and may instead indicate some residual systematic in the reconstructed CMB temperature maps. While tremendous effort has been focused on the removal of all known CMB foregrounds \citep[see e.g.][]{Sureau2014}, residual contamination coming from unresolved extragalactic point sources might still contaminate the ISW measurements and cosmological parameter estimation \citep{Serra2008,Millea2012}. \cite{ho} discussed how dust from galaxies at all redshifts contributes to the CMB temperature fluctuations, which, in turn, would result in a positive correlation between CMB temperatures and galaxy density. \cite{Yershov2012} also detected unexpected correlations between supernova redshifts and CMB temperatures. The same authors also analyzed {\it Planck} data and concluded that SN Ia measurements especially show this curious correlation \citep{Yershov2014}. Therefore, it is possible that the CMB data currently used for cosmological constraints might be affected by some remnant contamination that can affect our ISW measurements as well.

On the other hand, the excess ISW-like signals might indicate {\it new} physical effects at the largest scales. \cite{Nadathur2012} raised the possibility that non-Gaussianities in the primordial perturbations might be related to the excess ISW signals but this possibility appears to be excluded by recent {\it Planck} constraints \citep{Planck2015NonGauss}. Modified gravity theories with alternative growth rates, however, might provide some ground to discuss such excess signals.

Further analyses of \redmagic\ galaxies in the full Dark Energy Survey footprint, and synergy with the analysis of galaxy ``troughs" \citep{Gruen2016} and mass maps \citep{Vikram2015} will provide even more numerous catalogues of voids and superclusters to look for similar signals. In the near future, a factor of $\sim1/\sqrt{5}$ smaller error bars are expected for the full 5000 $deg^{2}$ DES Y5 footprint. We will impose more precise constraints on the ISW-like imprint of super-structures, including extra tests of the largest voids in the catalogue that might be responsible for any excess signal.

Advanced matched-filtering techniques introduced by \cite{NadathurCrittenden2016} and upcoming spectroscopic surveys \citep[e.g. DESI,][]{DESI} will also increase the sensitivity of stacking methods to decide if there is a real excess ISW(-like) signal, or if the patterns found in the SDSS and DES super-structure data sets are chance fluctuations.

\section*{Acknowledgments}
Funding for this project was partially provided by the Spanish Ministerio de Econom\'ia y Competitividad (MINECO) under projects FPA2012-39684, and Centro de Excelencia Severo Ochoa SEV-2012-0234 and SEV-2012-0249. 

We thank Eduardo Rozo, Eli Rykoff, and Risa Wechsler for creating observed and simulated DES \redmagic\ catalogues. We are also grateful for the insightful comments by Joseph Clampitt and Bhuvnesh Jain. 

The authors thank the Jubilee team for providing their LRG mock data and ISW maps based on their N-body simulation that was performed on the Juropa supercomputer of the J\"ulich Supercomputing Centre (JSC).

This paper has gone through internal review by the DES collaboration. It has been assigned DES paper id DES-2016-0178 and IFT preprint number IFT-UAM/CSIC-16-089, and Fermilab preprint number FERMILAB-PUB-16-448.

Funding for the DES Projects has been provided by the U.S. Department of Energy, the U.S. National Science Foundation, the Ministry of Science and Education of Spain, 
the Science and Technology Facilities Council of the United Kingdom, the Higher Education Funding Council for England, the National Center for Supercomputing 
Applications at the University of Illinois at Urbana-Champaign, the Kavli Institute of Cosmological Physics at the University of Chicago, 
the Center for Cosmology and Astro-Particle Physics at the Ohio State University,
the Mitchell Institute for Fundamental Physics and Astronomy at Texas A\&M University, Financiadora de Estudos e Projetos, 
Funda{\c c}{\~a}o Carlos Chagas Filho de Amparo {\`a} Pesquisa do Estado do Rio de Janeiro, Conselho Nacional de Desenvolvimento Cient{\'i}fico e Tecnol{\'o}gico and 
the Minist{\'e}rio da Ci{\^e}ncia, Tecnologia e Inova{\c c}{\~a}o, the Deutsche Forschungsgemeinschaft and the Collaborating Institutions in the Dark Energy Survey. 

The Collaborating Institutions are Argonne National Laboratory, the University of California at Santa Cruz, the University of Cambridge, Centro de Investigaciones Energ{\'e}ticas, 
Medioambientales y Tecnol{\'o}gicas-Madrid, the University of Chicago, University College London, the DES-Brazil Consortium, the University of Edinburgh, 
the Eidgen{\"o}ssische Technische Hochschule (ETH) Z{\"u}rich, 
Fermi National Accelerator Laboratory, the University of Illinois at Urbana-Champaign, the Institut de Ci{\`e}ncies de l'Espai (IEEC/CSIC), 
the Institut de F{\'i}sica d'Altes Energies, Lawrence Berkeley National Laboratory, the Ludwig-Maximilians Universit{\"a}t M{\"u}nchen and the associated Excellence Cluster Universe, 
the University of Michigan, the National Optical Astronomy Observatory, the University of Nottingham, The Ohio State University, the University of Pennsylvania, the University of Portsmouth, 
SLAC National Accelerator Laboratory, Stanford University, the University of Sussex, Texas A\&M University, and the OzDES Membership Consortium.

The DES data management system is supported by the National Science Foundation under Grant Number AST-1138766.
The DES participants from Spanish institutions are partially supported by MINECO under grants AYA2012-39559, ESP2013-48274, FPA2013-47986, and Centro de Excelencia Severo Ochoa SEV-2012-0234 and SEV-2012-0249.
Research leading to these results has received funding from the European Research Council under the European Union's Seventh Framework Programme (FP7/2007-2013) including ERC grant agreements 
 240672, 291329, and 306478. Support for DG was provided by NASA through the Einstein Fellowship Program, grant PF5-160138.

We are grateful for the extraordinary contributions of our CTIO colleagues and the DECam Construction, Commissioning and Science Verification teams in achieving the excellent instrument and telescope conditions that have made this work possible. The success of this project also relies critically on the expertise and dedication of the DES Data Management group.

\bibliographystyle{mn2e}
\bibliography{refs}

\begin{thebibliography}{}

\bibitem[\protect\citeauthoryear{{Aiola}, {Kosowsky} \& {Wang}}{{Aiola}
  et~al.}{2015}]{Aiola}
{Aiola} S.,  {Kosowsky} A.,    {Wang} B.,  2015, \prd, 91, 043510

\bibitem[\protect\citeauthoryear{{Boughn} \& {Crittenden}}{{Boughn} \&
  {Crittenden}}{2004}]{b11}
{Boughn} S.,  {Crittenden} R.,  2004, Nature, 427, 45

\bibitem[\protect\citeauthoryear{{Bremer}, {Silk}, {Davies} \&
  {Lehnert}}{{Bremer} et~al.}{2010}]{BremerEtal2010}
{Bremer} M.~N.,  {Silk} J.,  {Davies} L.~J.~M.,    {Lehnert} M.~D.,  2010,
  \mnras, 404, L69

\bibitem[\protect\citeauthoryear{{Cai}, {Cole}, {Jenkins} \& {Frenk}}{{Cai}
  et~al.}{2010}]{CaiEtAl2010}
{Cai} Y.-C.,  {Cole} S.,  {Jenkins} A.,    {Frenk} C.~S.,  2010, \mnras, 407,
  201

\bibitem[\protect\citeauthoryear{{Cai}, {Li}, {Cole}, {Frenk} \&
  {Neyrinck}}{{Cai} et~al.}{2014}]{CaiEtAl2014}
{Cai} Y.-C.,  {Li} B.,  {Cole} S.,  {Frenk} C.~S.,    {Neyrinck} M.,  2014,
  \mnras, 439, 2978

\bibitem[\protect\citeauthoryear{{Cai}, {Neyrinck}, {Mao}, {Peacock}, {Szapudi}
  \& {Berlind}}{{Cai} et~al.}{2016}]{Cai2016}
{Cai} Y.-C.,  {Neyrinck} M.,  {Mao} Q.,  {Peacock} J.~A.,  {Szapudi} I.,
  {Berlind} A.~A.,  2016, ArXiv e-prints: 1609.00301

\bibitem[\protect\citeauthoryear{{Cai}, {Neyrinck}, {Szapudi}, {Cole} \&
  {Frenk}}{{Cai} et~al.}{2014}]{CaiEtAl2013}
{Cai} Y.-C.,  {Neyrinck} M.~C.,  {Szapudi} I.,  {Cole} S.,    {Frenk} C.~S.,
  2014, \apj, 786, 110

\bibitem[\protect\citeauthoryear{{Clampitt} \& {Jain}}{{Clampitt} \&
  {Jain}}{2015}]{ClampittJain2015}
{Clampitt} J.,  {Jain} B.,  2015, \mnras, 454, 3357

\bibitem[\protect\citeauthoryear{{Collister}, {Lahav}, {Blake}, {Cannon},
  {Croom}, {Drinkwater}, {Edge}, {Eisenstein}, {Loveday}, {Nichol}, {Pimbblet},
  {de Propris}, {Roseboom}, {Ross}, {Schneider}, {Shanks} \&
  {Wake}}{{Collister} et~al.}{2007}]{megaz}
{Collister} A.,  {Lahav} O.,  {Blake} C.,  {Cannon} R.,  {Croom} S.,
  {Drinkwater} M.,  {Edge} A.,  {Eisenstein} D.,  {Loveday} J.,  {Nichol} R.,
  {Pimbblet} K.,  {de Propris} R.,  {Roseboom} I.,  {Ross} N.,  {Schneider}
  D.~P.,  {Shanks} T.,    {Wake} D.,  2007, \mnras, 375, 68

\bibitem[\protect\citeauthoryear{{Dark Energy Survey Collaboration}}{{Dark
  Energy Survey Collaboration}}{2016}]{morethanDE2016}
{Dark Energy Survey Collaboration} 2016, \mnras, 460, 1270

\bibitem[\protect\citeauthoryear{{Einasto}, {Suhhonenko}, {H{\"u}tsi}, {Saar},
  {Einasto}, {Liivam{\"a}gi}, {M{\"u}ller}, {Starobinsky}, {Tago} \&
  {Tempel}}{{Einasto} et~al.}{2011}]{Einasto2011}
{Einasto} J.,  {Suhhonenko} I.,  {H{\"u}tsi} G.,  {Saar} E.,  {Einasto} M.,
  {Liivam{\"a}gi} L.~J.,  {M{\"u}ller} V.,  {Starobinsky} A.~A.,  {Tago} E.,
  {Tempel} E.,  2011, \aap, 534, A128

\bibitem[\protect\citeauthoryear{{Eisenstein}, {Zehavi}, {Hogg} \& et
  al.}{{Eisenstein} et~al.}{2005}]{Eisenstein2005}
{Eisenstein} D.~J.,  {Zehavi} I.,  {Hogg} D.~W.,    et al. 2005, \apj, 633, 560

\bibitem[\protect\citeauthoryear{{Finelli}, {Garc{\'{\i}}a-Bellido},
  {Kov{\'a}cs}, {Paci} \& {Szapudi}}{{Finelli} et~al.}{2015}]{FinelliEtal2014}
{Finelli} F.,  {Garc{\'{\i}}a-Bellido} J.,  {Kov{\'a}cs} A.,  {Paci} F.,
  {Szapudi} I.,  2015, \mnras, 455, 1246

\bibitem[\protect\citeauthoryear{{Flaugher}, {Diehl}, {Honscheid} \& {The DES
  Collaboration}}{{Flaugher} et~al.}{2015}]{DECam}
{Flaugher} B.,  {Diehl} H.~T.,  {Honscheid} K.,    {The DES Collaboration}
  2015, \aj, 150, 150

\bibitem[\protect\citeauthoryear{{Flender}, {Hotchkiss} \&
  {Nadathur}}{{Flender} et~al.}{2013}]{Flender2013}
{Flender} S.,  {Hotchkiss} S.,    {Nadathur} S.,  2013, JCAP, 2, 13

\bibitem[\protect\citeauthoryear{{Fosalba} \& {Gazta{\~n}aga}}{{Fosalba} \&
  {Gazta{\~n}aga}}{2004}]{Fosalba2004}
{Fosalba} P.,  {Gazta{\~n}aga} E.,  2004, \mnras, 350, L37

\bibitem[\protect\citeauthoryear{{Fosalba}, {Gazta{\~n}aga} \&
  {Castander}}{{Fosalba} et~al.}{2003}]{Fosalba2003}
{Fosalba} P.,  {Gazta{\~n}aga} E.,    {Castander} F.~J.,  2003, \apjl, 597, L89

\bibitem[\protect\citeauthoryear{{Francis} \& {Peacock}}{{Francis} \&
  {Peacock}}{2010}]{b14}
{Francis} C.~L.,  {Peacock} J.~A.,  2010, MNRAS, 406, 2

\bibitem[\protect\citeauthoryear{{Giannantonio}, {Crittenden}, {Nichol} \&
  {Ross}}{{Giannantonio} et~al.}{2012}]{gian}
{Giannantonio} T.,  {Crittenden} R.,  {Nichol} R.,    {Ross} A.~J.,  2012,
  \mnras, 426, 2581

\bibitem[\protect\citeauthoryear{{Giannantonio}, {Scranton}, {Crittenden} \& et
  al.}{{Giannantonio} et~al.}{2008}]{gianEtAl08}
{Giannantonio} T.,  {Scranton} R.,  {Crittenden} R.~G.,    et al. 2008, \prd,
  77, 123520

\bibitem[\protect\citeauthoryear{{Gorski}, {Hivon} \& et al.}{{Gorski}
  et~al.}{2005}]{healpix}
{Gorski} K.~M.,  {Hivon} E.,    et al. 2005, \apj, 622, 759

\bibitem[\protect\citeauthoryear{{Granett}, {Kov{\'a}cs} \& {Hawken}}{{Granett}
  et~al.}{2015}]{Granett2015}
{Granett} B.~R.,  {Kov{\'a}cs} A.,    {Hawken} A.~J.,  2015, \mnras, 454, 2804

\bibitem[\protect\citeauthoryear{{Granett}, {Neyrinck} \& {Szapudi}}{{Granett}
  et~al.}{2008}]{GranettEtAl2008}
{Granett} B.~R.,  {Neyrinck} M.~C.,    {Szapudi} I.,  2008, \apjl, 683, L99

\bibitem[\protect\citeauthoryear{{Gruen}, {Friedrich}, {Amara}, {Bacon},
  {Bonnett} \& et al.}{{Gruen} et~al.}{2016}]{Gruen2016}
{Gruen} D.,  {Friedrich} O.,  {Amara} A.,  {Bacon} D.,  {Bonnett} C.,    et al.
  2016, \mnras, 455, 3367

\bibitem[\protect\citeauthoryear{{Hern{\'a}ndez-Monteagudo} \&
  {Smith}}{{Hern{\'a}ndez-Monteagudo} \& {Smith}}{2013}]{Hernandez2013}
{Hern{\'a}ndez-Monteagudo} C.,  {Smith} R.~E.,  2013, \mnras, 435, 1094

\bibitem[\protect\citeauthoryear{{Hinshaw}, {Larson}, {Komatsu} \& {et
  al.}}{{Hinshaw} et~al.}{2013}]{WMAP9}
{Hinshaw} G.,  {Larson} D.,  {Komatsu} E.,    {et al.} 2013, \apjs, 208, 19

\bibitem[\protect\citeauthoryear{{Ho}, {Hirata}, {Padmanabhan}, {Seljak} \&
  {Bahcall}}{{Ho} et~al.}{2008}]{ho}
{Ho} S.,  {Hirata} C.,  {Padmanabhan} N.,  {Seljak} U.,    {Bahcall} N.,  2008,
  Physical Review D, 78, 043519

\bibitem[\protect\citeauthoryear{{Hotchkiss}, {Nadathur}, {Gottl{\"o}ber},
  {Iliev}, {Knebe}, {Watson} \& {Yepes}}{{Hotchkiss}
  et~al.}{2015}]{Hotchkiss2015}
{Hotchkiss} S.,  {Nadathur} S.,  {Gottl{\"o}ber} S.,  {Iliev} I.~T.,  {Knebe}
  A.,  {Watson} W.~A.,    {Yepes} G.,  2015, \mnras, 446, 1321

\bibitem[\protect\citeauthoryear{{Ili{\'c}}, {Langer} \& {Douspis}}{{Ili{\'c}}
  et~al.}{2013}]{Ilic2013}
{Ili{\'c}} S.,  {Langer} M.,    {Douspis} M.,  2013, \aap, 556, A51

\bibitem[\protect\citeauthoryear{{Ili{\'c}}, {Langer} \& {Douspis}}{{Ili{\'c}}
  et~al.}{2014}]{Ilic_ERRATUM}
{Ili{\'c}} S.,  {Langer} M.,    {Douspis} M.,  2014, \aap, 572, C2

\bibitem[\protect\citeauthoryear{{Kazin}, {Blanton}, {Scoccimarro}, {McBride},
  {Berlind}, {Bahcall}, {Brinkmann}, {Czarapata}, {Frieman}, {Kent},
  {Schneider} \& {Szalay}}{{Kazin} et~al.}{2010}]{Kazin2010}
{Kazin} E.~A.,  {Blanton} M.~R.,  {Scoccimarro} R.,  {McBride} C.~K.,
  {Berlind} A.~A.,  {Bahcall} N.~A.,  {Brinkmann} J.,  {Czarapata} P.,
  {Frieman} J.~A.,  {Kent} S.~M.,  {Schneider} D.~P.,    {Szalay} A.~S.,  2010,
  \apj, 710, 1444

\bibitem[\protect\citeauthoryear{{Kov{\'a}cs} \&
  {Garc{\'{\i}}a-Bellido}}{{Kov{\'a}cs} \&
  {Garc{\'{\i}}a-Bellido}}{2016}]{KovacsJGB2015}
{Kov{\'a}cs} A.,  {Garc{\'{\i}}a-Bellido} J.,  2016, \mnras, 462, 1882

\bibitem[\protect\citeauthoryear{{Kov{\'a}cs} \& {Granett}}{{Kov{\'a}cs} \&
  {Granett}}{2015}]{Kovacs2015}
{Kov{\'a}cs} A.,  {Granett} B.~R.,  2015, \mnras, 452, 1295

\bibitem[\protect\citeauthoryear{{Kov{\'a}cs}, {Szapudi}, {Granett} \&
  {Frei}}{{Kov{\'a}cs} et~al.}{2013}]{KovacsEtAl2013}
{Kov{\'a}cs} A.,  {Szapudi} I.,  {Granett} B.~R.,    {Frei} Z.,  2013, \mnras,
  431, L28

\bibitem[\protect\citeauthoryear{{Levi}, {Bebek}, {Beers}, {Blum}, {Cahn},
  {Eisenstein}, {Flaugher}, {Honscheid}, {Kron}, {Lahav}, {McDonald}, {Roe},
  {Schlegel} \& {representing the DESI collaboration}}{{Levi}
  et~al.}{2013}]{DESI}
{Levi} M.,  {Bebek} C.,  {Beers} T.,  {Blum} R.,  {Cahn} R.,  {Eisenstein} D.,
  {Flaugher} B.,  {Honscheid} K.,  {Kron} R.,  {Lahav} O.,  {McDonald} P.,
  {Roe} N.,  {Schlegel} D.,    {representing the DESI collaboration} 2013,
  ArXiv e-prints: 1308.0847

\bibitem[\protect\citeauthoryear{{Marcos-Caballero}, {Fern{\'a}ndez-Cobos},
  {Mart{\'{\i}}nez-Gonz{\'a}lez} \& {Vielva}}{{Marcos-Caballero}
  et~al.}{2015}]{MarcosCaballero2015}
{Marcos-Caballero} A.,  {Fern{\'a}ndez-Cobos} R.,
  {Mart{\'{\i}}nez-Gonz{\'a}lez} E.,    {Vielva} P.,  2015, ArXiv e-prints

\bibitem[\protect\citeauthoryear{{Millea}, {Dor{\'e}}, {Dudley}, {Holder},
  {Knox}, {Shaw}, {Song} \& {Zahn}}{{Millea} et~al.}{2012}]{Millea2012}
{Millea} M.,  {Dor{\'e}} O.,  {Dudley} J.,  {Holder} G.,  {Knox} L.,  {Shaw}
  L.,  {Song} Y.-S.,    {Zahn} O.,  2012, \apj, 746, 4

\bibitem[\protect\citeauthoryear{{Nadathur}}{{Nadathur}}{2016}]{Nadathur2016}
{Nadathur} S.,  2016, \mnras, 461, 358

\bibitem[\protect\citeauthoryear{{Nadathur} \& {Crittenden}}{{Nadathur} \&
  {Crittenden}}{2016}]{NadathurCrittenden2016}
{Nadathur} S.,  {Crittenden} R.,  2016, ArXiv e-prints: 1608.08638

\bibitem[\protect\citeauthoryear{{Nadathur} \& {Hotchkiss}}{{Nadathur} \&
  {Hotchkiss}}{2015}]{Nadathur2015}
{Nadathur} S.,  {Hotchkiss} S.,  2015, \mnras, 454, 889

\bibitem[\protect\citeauthoryear{{Nadathur}, {Hotchkiss} \&
  {Sarkar}}{{Nadathur} et~al.}{2012}]{Nadathur2012}
{Nadathur} S.,  {Hotchkiss} S.,    {Sarkar} S.,  2012, JCAP, 6, 42

\bibitem[\protect\citeauthoryear{{Nadathur}, {Lavinto}, {Hotchkiss} \&
  {R{\"a}s{\"a}nen}}{{Nadathur} et~al.}{2014}]{Nadathur2014}
{Nadathur} S.,  {Lavinto} M.,  {Hotchkiss} S.,    {R{\"a}s{\"a}nen} S.,  2014,
  \prd, 90, 103510

\bibitem[\protect\citeauthoryear{{Naidoo}, {Benoit-L{\'e}vy} \&
  {Lahav}}{{Naidoo} et~al.}{2016}]{Naidoo2016}
{Naidoo} K.,  {Benoit-L{\'e}vy} A.,    {Lahav} O.,  2016, \mnras, 459, L71

\bibitem[\protect\citeauthoryear{{Neyrinck}}{{Neyrinck}}{2008}]{ZOBOV}
{Neyrinck} M.~C.,  2008, \mnras, 386, 2101

\bibitem[\protect\citeauthoryear{{P{\'a}pai} \& {Szapudi}}{{P{\'a}pai} \&
  {Szapudi}}{2010}]{PapaiSzapudi2010}
{P{\'a}pai} P.,  {Szapudi} I.,  2010, \apj, 725, 2078

\bibitem[\protect\citeauthoryear{{P{\'a}pai}, {Szapudi} \&
  {Granett}}{{P{\'a}pai} et~al.}{2011}]{PapaiEtAl2011}
{P{\'a}pai} P.,  {Szapudi} I.,    {Granett} B.~R.,  2011, \apj, 732, 27

\bibitem[\protect\citeauthoryear{{Planck 2013 results. XIX.}}{{Planck 2013
  results. XIX.}}{2014}]{Planck19}
{Planck 2013 results. XIX.} 2014, \aap, 571, A19

\bibitem[\protect\citeauthoryear{{Planck 2015 results. XI.}}{{Planck 2015
  results. XI.}}{2015}]{Planck_15}
{Planck 2015 results. XI.} 2015, ArXiv e-prints: 1507.02704

\bibitem[\protect\citeauthoryear{{Planck 2015 results. XVII.}}{{Planck 2015
  results. XVII.}}{2015}]{Planck2015NonGauss}
{Planck 2015 results. XVII.} 2015, ArXiv e-prints: 1502.01592

\bibitem[\protect\citeauthoryear{{Planck 2015 results. XXI.}}{{Planck 2015
  results. XXI.}}{2015}]{PlackISW2015}
{Planck 2015 results. XXI.} 2015, ArXiv e-prints: 1502.01595

\bibitem[\protect\citeauthoryear{{Rees} \& {Sciama}}{{Rees} \&
  {Sciama}}{1968}]{ReesSciama}
{Rees} M.~J.,  {Sciama} D.~W.,  1968, \nat, 217, 511

\bibitem[\protect\citeauthoryear{{Rozo}, {Rykoff}, {Abate} \& et al.}{{Rozo}
  et~al.}{2016}]{Rozo2015}
{Rozo} E.,  {Rykoff} E.~S.,  {Abate} A.,    et al. 2016, \mnras, 461, 1431

\bibitem[\protect\citeauthoryear{{Rykoff}, {Rozo}, {Busha} \& et al.}{{Rykoff}
  et~al.}{2014}]{Rykoff2014}
{Rykoff} E.~S.,  {Rozo} E.,  {Busha} M.~T.,    et al. 2014, \apj, 785, 104

\bibitem[\protect\citeauthoryear{{Sachs} \& {Wolfe}}{{Sachs} \&
  {Wolfe}}{1967}]{SachsWolfe}
{Sachs} R.~K.,  {Wolfe} A.~M.,  1967, ApJL, 147, 73

\bibitem[\protect\citeauthoryear{{Sahlen}, {Zubeldia} \& {Silk}}{{Sahlen}
  et~al.}{2015}]{Sahlen2015}
{Sahlen} M.,  {Zubeldia} I.,    {Silk} J.,  2015, ArXiv e-prints

\bibitem[\protect\citeauthoryear{{S{\'a}nchez}, {Carrasco Kind}, {Lin} \& et
  al.}{{S{\'a}nchez} et~al.}{2014}]{Sanchez2014}
{S{\'a}nchez} C.,  {Carrasco Kind} M.,  {Lin} H.,    et al. 2014, \mnras, 445,
  1482

\bibitem[\protect\citeauthoryear{{S{\'a}nchez}, {Clampitt}, {Kovacs} \& et
  al.}{{S{\'a}nchez} et~al.}{2016}]{Sanchez2016}
{S{\'a}nchez} C.,  {Clampitt} J.,  {Kovacs} A.,    et al. 2016, ArXiv e-prints:
  1605.03982

\bibitem[\protect\citeauthoryear{{Serra}, {Cooray}, {Amblard}, {Pagano} \&
  {Melchiorri}}{{Serra} et~al.}{2008}]{Serra2008}
{Serra} P.,  {Cooray} A.,  {Amblard} A.,  {Pagano} L.,    {Melchiorri} A.,
  2008, \prd, 78, 043004

\bibitem[\protect\citeauthoryear{{Sureau}, {Starck}, {Bobin}, {Paykari} \&
  {Rassat}}{{Sureau} et~al.}{2014}]{Sureau2014}
{Sureau} F.~C.,  {Starck} J.-L.,  {Bobin} J.,  {Paykari} P.,    {Rassat} A.,
  2014, \aap, 566, A100

\bibitem[\protect\citeauthoryear{{Sutter}, {Lavaux}, {Wandelt}, {Weinberg} \&
  {Warren}}{{Sutter} et~al.}{2014}]{Sutter2014_DM}
{Sutter} P.~M.,  {Lavaux} G.,  {Wandelt} B.~D.,  {Weinberg} D.~H.,    {Warren}
  M.~S.,  2014, \mnras, 438, 3177

\bibitem[\protect\citeauthoryear{{Szapudi}, {Kov{\'a}cs}, {Granett}, {Frei},
  {Silk}, {Burgett}, {Cole}, {Draper}, {Farrow}, {Kaiser}, {Magnier},
  {Metcalfe}, {Morgan}, {Price}, {Tonry} \& {Wainscoat}}{{Szapudi}
  et~al.}{2015}]{SzapudiEtAl2014}
{Szapudi} I.,  {Kov{\'a}cs} A.,  {Granett} B.~R.,  {Frei} Z.,  {Silk} J.,
  {Burgett} W.,  {Cole} S.,  {Draper} P.~W.,  {Farrow} D.~J.,  {Kaiser} N.,
  {Magnier} E.~A.,  {Metcalfe} N.,  {Morgan} J.~S.,  {Price} P.,  {Tonry} J.,
   {Wainscoat} R.,  2015, \mnras, 450, 288

\bibitem[\protect\citeauthoryear{{The Dark Energy Survey Collaboration}}{{The
  Dark Energy Survey Collaboration}}{2005}]{DES}
{The Dark Energy Survey Collaboration} 2005, arXiv:astro-ph/0510346

\bibitem[\protect\citeauthoryear{{Vikram}, {Chang}, {Jain}, {Bacon}, {Amara},
  {Becker} \& et al.}{{Vikram} et~al.}{2015}]{Vikram2015}
{Vikram} V.,  {Chang} C.,  {Jain} B.,  {Bacon} D.,  {Amara} A.,  {Becker}
  M.~R.,    et al. 2015, \prd, 92, 022006

\bibitem[\protect\citeauthoryear{{Watson}, {Diego}, {Gottl{\"o}ber}, {Iliev},
  {Knebe}, {Mart{\'{\i}}nez-Gonz{\'a}lez}, {Yepes}, {Barreiro},
  {Gonz{\'a}lez-Nuevo}, {Hotchkiss}, {Marcos-Caballero}, {Nadathur} \&
  {Vielva}}{{Watson} et~al.}{2014}]{Watson2014}
{Watson} W.~A.,  {Diego} J.~M.,  {Gottl{\"o}ber} S.,  {Iliev} I.~T.,  {Knebe}
  A.,  {Mart{\'{\i}}nez-Gonz{\'a}lez} E.,  {Yepes} G.,  {Barreiro} R.~B.,
  {Gonz{\'a}lez-Nuevo} J.,  {Hotchkiss} S.,  {Marcos-Caballero} A.,  {Nadathur}
  S.,    {Vielva} P.,  2014, \mnras, 438, 412

\bibitem[\protect\citeauthoryear{{Yershov}, {Orlov} \& {Raikov}}{{Yershov}
  et~al.}{2012}]{Yershov2012}
{Yershov} V.~N.,  {Orlov} V.~V.,    {Raikov} A.~A.,  2012, \mnras, 423, 2147

\bibitem[\protect\citeauthoryear{{Yershov}, {Orlov} \& {Raikov}}{{Yershov}
  et~al.}{2014}]{Yershov2014}
{Yershov} V.~N.,  {Orlov} V.~V.,    {Raikov} A.~A.,  2014, \mnras, 445, 2440

\end{thebibliography}

\section*{Affiliations}
$^{1}$ Institut de F\'{\i}sica d'Altes Energies (IFAE), The Barcelona Institute of Science and Technology, Campus UAB, 08193 Bellaterra (Barcelona), Spain\\
$^{2}$ Instituto de F\'isica Te\'orica IFT-UAM/CSIC, Universidad Aut\'onoma de Madrid, Cantoblanco 28049 Madrid, Spain\\
$^{3}$ Institute of Cosmology \& Gravitation, University of Portsmouth, Portsmouth, PO1 3FX, UK\\
$^{4}$ Kavli Institute for Particle Astrophysics \& Cosmology, P. O. Box 2450, Stanford University, Stanford, CA 94305, USA\\
$^{5}$ SLAC National Accelerator Laboratory, Menlo Park, CA 94025, USA\\
$^{6}$ Einstein Fellow\\
$^{7}$ Department of Physics, University of Michigan, Ann Arbor, MI 48109, USA\\
$^{8}$ Department of Physics, Stanford University, 382 Via Pueblo Mall, Stanford, CA 94305, USA\\
$^{9}$ Fermi National Accelerator Laboratory, P. O. Box 500, Batavia, IL 60510, USA\\
$^{10}$ Kavli Institute for Cosmological Physics, University of Chicago, Chicago, IL 60637, USA\\
$^{11}$ Institut de Ci\`encies de l'Espai, IEEC-CSIC, Campus UAB, Carrer de Can Magrans, s/n,  08193 Bellaterra, Barcelona, Spain\\
$^{12}$ Department of Physics \& Astronomy, University College London, Gower Street, London, WC1E 6BT, UK\\
$^{13}$ Instituci\'o Catalana de Recerca i Estudis Avan\c{c}ats, E-08010 Barcelona, Spain\\
$^{14}$ Institute of Astronomy, University of Cambridge, Mading- ley Road, Cambridge CB3 0HA, UK\\
$^{15}$ Kavli Institute for Cosmology, University of Cambridge, Madingley Road, Cambridge CB3 0HA, UK\\
$^{16}$ Department of Physics and Electronics, Rhodes University, PO Box 94, Grahamstown, 6140, South Africa\\
$^{17}$ CNRS, UMR 7095, Institut d'Astrophysique de Paris, F-75014, Paris, France\\
$^{18}$ Sorbonne Universit\'es, UPMC Univ Paris 06, UMR 7095, Institut d'Astrophysique de Paris, F-75014, Paris, France\\
$^{19}$ Laborat\'orio Interinstitucional de e-Astronomia - LIneA, Rua Gal. Jos\'e Cristino 77, Rio de Janeiro, RJ - 20921-400, Brazil\\
$^{20}$ Observat\'orio Nacional, Rua Gal. Jos\'e Cristino 77, Rio de Janeiro, RJ - 20921-400, Brazil\\
$^{21}$ Department of Astronomy, University of Illinois, 1002 W. Green Street, Urbana, IL 61801, USA\\
$^{22}$ National Center for Supercomputing Applications, 1205 West Clark St., Urbana, IL 61801, USA\\
$^{23}$ School of Physics and Astronomy, University of Southampton,  Southampton, SO17 1BJ, UK\\
$^{24}$ George P. and Cynthia Woods Mitchell Institute for Fundamental Physics and Astronomy, and Department of Physics and Astronomy, Texas A\&M University, College Station, TX 77843,  USA\\
$^{25}$ Excellence Cluster Universe, Boltzmannstr.\ 2, 85748 Garching, Germany\\
$^{26}$ Faculty of Physics, Ludwig-Maximilians-Universit\"at, Scheinerstr. 1, 81679 Munich, Germany\\
$^{27}$ Jet Propulsion Laboratory, California Institute of Technology, 4800 Oak Grove Dr., Pasadena, CA 91109, USA\\
$^{28}$ Department of Astronomy, University of California, Berkeley, 501 Campbell Hall 3411, Berkeley, CA 94720\\
$^{29}$ Lawrence Berkeley National Laboratory, 1 Cyclotron Road, Berkeley, CA 94720, USA\\
$^{30}$ Cerro Tololo Inter-American Observatory, National Optical Astronomy Observatory, Casilla 603, La Serena, Chile\\
$^{31}$ Australian Astronomical Observatory, North Ryde, NSW 2113, Australia\\
$^{32}$ Carnegie Observatories, 813 Santa Barbara St., Pasadena, CA 91101, USA\\
$^{33}$ Department of Astrophysical Sciences, Princeton University, Peyton Hall, Princeton, NJ 08544, USA\\
$^{34}$ Department of Physics and Astronomy, Pevensey Building, University of Sussex, Brighton, BN1 9QH, UK\\
$^{35}$ Centro de Investigaciones Energ\'eticas, Medioambientales y Tecnol\'ogicas (CIEMAT), Madrid, Spain\\
$^{36}$ ICTP South American Institute for Fundamental Research\\ Instituto de F\'{\i}sica Te\'orica, Universidade Estadual Paulista, S\~ao Paulo, Brazil\\
$^{37}$ Department of Physics and Astronomy, University of Pennsylvania, Philadelphia, PA 19104, USA\\
$^{38}$ Institute for Astronomy, University of Edinburgh, Royal Observatory, Blackford Hill, Edinburgh, EH9 3HJ, UK\\
\end{document}